\documentclass{article}
\usepackage{amsfonts}
\usepackage{amssymb}
\usepackage{graphicx}
\usepackage{amsmath}
\usepackage{amsthm}

\nonstopmode
\newtheorem{theorem}{Theorem}[section]
\newtheorem{proposition}[theorem]{Proposition}
\newtheorem{lemma}[theorem]{Lemma}

\newtheorem{corollary}[theorem]{Corollary}

\theoremstyle{definition}
\newtheorem{definition}[theorem]{Definition}

\theoremstyle{remark}
\newtheorem{example}[theorem]{Example}
\newtheorem{remark}[theorem]{Remark}
\begin{document}

\title{Indifference price with general semimartingales }
\author{ Sara Biagini\footnotemark[1] \  \and Marco Frittelli 
\footnotemark[4] \  \and Matheus Grasselli \footnotemark[3] }
\date{April 18, 2008}
\maketitle

\begin{abstract}
For utility functions $u$ finite valued on $\mathbb{R}$, we prove a duality
formula for utility maximization with random endowment in general
semimartingale incomplete markets. The main novelty of the paper is that
possibly non locally bounded semimartingale price processes are allowed.
Following Biagini and Frittelli \cite{BiaFri06}, the analysis is based on
the duality between the Orlicz spaces $(L^{\widehat{u}}, (L^{\widehat{u}})^*
)$ naturally associated to the utility function. This formulation enables
several key properties of the indifference price $\pi(B)$ of a claim $B$
satisfying conditions weaker than those assumed in literature. In
particular, the indifference price functional $\pi$ turns out to be, apart
from a sign, a convex risk measure on the Orlicz space $L^{\widehat{u}}$.
\end{abstract}

\renewcommand{\thefootnote}{\fnsymbol{footnote}} \footnotetext[1]{%
Universit\`{a} di Pisa, Italy. Email: sara.biagini@ec.unipi.it} 
\footnotetext[4]{%
Universit\`{a} di Milano, Italy. Email: marco.frittelli@mat.unimi.it} 
\footnotetext[3]{%
McMaster University, Canada. Email: grasselli@math.mcmaster.ca}

\renewcommand{\thefootnote}{\arabic{footnote}}

\noindent \textbf{Key words:} Indifference price - utility maximization --
non locally bounded semimartingale -- random endowment - incomplete market
-- Orlicz space -- convex duality - convex risk measure

\noindent \textbf{JEL Classification:} G11, G12, G13\qquad \newline
\textbf{Mathematics Subject Classification (2000):} primary 60G48, 60G44,
49N15, 91B28; secondary 46E30, 46N30, 91B16.

\section{Introduction}

\label{introduction}

The main purpose of this paper is to study the indifference pricing
framework in markets where the underlying traded assets are described by
general semimartingales which are \emph{not assumed to be locally bounded}.
Following Hodges and~Neuberger \cite{hn}, we define the \emph{(seller)
indifference price} $\pi (B)$ of a claim $B$ as the implicit solution of the
equation

\begin{equation}
\sup_{H\in \mathcal{H}^{W}}E\left[ u\left( x+\int_{0}^{T}H_tdS_t\right) %
\right] =\sup_{H\in \mathcal{H}^{W}}E\left[ u\left( x+\pi
(B)+\int_{0}^{T}H_tdS_t-B\right) \right] ,  \label{def_indiff}
\end{equation}%
where $x\in \mathbb{R}$ is the constant initial endowment, $T<\infty $ is a
fixed time horizon while $S$ is an $\mathbb{R}^{d}-$valued c\`{a}dl\`{a}g
semimartingale defined on a filtered stochastic basis $(\Omega ,\mathcal{F},(%
\mathcal{F}_{t})_{t\in \lbrack 0,T]},P)$ that satisfies the usual
assumptions. The $\mathbb{R}^{d}-$valued portfolio process $H$ belongs to an
appropriate class $\mathcal{H}^{W}$ of admissible integrands defined in
Section 2.1 through a random variable $W$ that controls the losses incurred
in trading. $B$ is an $\mathcal{F}_{T}$--measurable random variable
corresponding to a financial liability at time $T$ and satisfies the
integrability conditions discussed in Section 3.1.

Throughout the paper, the utility function $u$ is assumed to be an
increasing and concave function $u:\mathbb{R}\rightarrow \mathbb{\mathbb{R}}$%
\ satisfying $\lim_{x\rightarrow -\infty }u(x)=-\infty .$\newline
\indent  Neither \emph{strict monotonicity nor strict concavity are required}%
, but we exclude that $u$ is constant on $\mathbb{R}$.

In principle, a general way to compute the indifference price in %
\eqref{def_indiff} is to solve the two utility maximization problems, in the
sense of finding the optimizers in the class of admissible integrands. Such
optimizers then correspond to the optimal trading strategies that an
investor should follow with or without the claim $B$, therefore providing a
corresponding notion of \emph{indifference hedging} for the claim. However,
it is generally possible to employ duality arguments to obtain the optimal 
\emph{values} for utility maximization problems under broader assumptions
than those necessary to find their optimizer. Since these values are all
that is necessary for calculating the indifference price itself, the main
goal here is the pursuit of such duality results rather than a full analysis
of the indifference hedging problem which is deferred to future work (even
though some partial results in this direction are provided in Proposition %
\ref{hedge}).

The key to establish such duality above is to choose convenient dual spaces
as the ambient for the domains of optimization. Our approach is to use the
Orlicz space $L^{\widehat{u}}$ - and its dual space $L^{\widehat{\Phi }}$ -
that arises naturally from the choice of the utility function $u$ and was
previously used in \cite{BiaFri06} for the special case of $B=0$, as
explained in Section 2.

We then use this general framework for the case of a random endowment $B$ in
Section 3 and prove in Theorem \ref{conj1} a duality result of the type 
\begin{eqnarray}
&&\sup_{H\in \mathcal{H}^{W}}E\left[ u\left( x+\int_{0}^{T}H_tdS_t-B\right) %
\right]  \label{1} \\
&=&\min_{\lambda >0,\text{ }Q\in \mathcal{M}^{W}}\left\{ \lambda x-\lambda
Q(B)+E\left[ \Phi \left( \lambda \frac{{dQ}^{r}}{{dP}}\right) \right]
+\lambda \Vert Q^{s}\Vert \right\} .  \label{2}
\end{eqnarray}%
where $W$ is a loss control and in the dual problem (\ref{2}), $\Phi :%
\mathbb{R}_{+}\rightarrow \mathbb{\mathbb{R}}$ is the convex conjugate of
the utility function $u$, defined by 
\begin{equation}
\Phi (y):=\sup_{x\in \mathbb{R}}\left\{ u(x)-xy\right\} ,  \label{phi_def}
\end{equation}%
while $\mathcal{M}^{W}$ is the appropriate set of linear pricing functionals 
$Q$, which admit the decomposition 
\begin{equation*}
Q=Q^{r}+Q^{s}
\end{equation*}%
into regular and singular parts. The penalty term in the right-hand side of (%
\ref{2}) is split into the expectation $E\left[ \Phi \left( \lambda \frac{{dQ%
}^{r}}{{dP}}\right) \right] $, associated only with the regular part of $Q$,
and the norm $\Vert Q^{s}\Vert $, associated only with its singular part.

>From the previous results \cite{BiaFri06} in the case $B=0$, we expected the
presence of the singular part $\Vert Q^{s}\Vert $, due to the fact that we
allow possibly unbounded semimartingales$.$ As shown in the Examples in
Section 3.6.1 and discussed in Section 3.5, when also the claim $B$ is
present and is not sufficiently integrable, in the above duality an
additional singular term appears from $Q(B)=E_{Q^{r}}[B]+Q^{s}(B).$

The above duality result (\ref{1})-(\ref{2}) holds under the assumptions
that $B$ belongs to the set $\mathcal{A}_{u}$ of \textit{admissible claims}
(see definition \ref{defA}). Even though we admit price processes
represented by general semimartingale, the above assumptions on $B$ are
weaker than those assumed in the literature for the locally bounded case -
see the discussion in Sections \ref{secClaim} and 5. This is a nice
consequence of the selection of the Orlicz space duality.

\bigskip

Regarding the primal utility maximization problem with random endowment, in
Theorem \ref{conj1} we also prove the existence of the optimal solution $%
f_{B}$ in a slightly enlarged set than $\{\int_{0}^{T}H_t dS_t \mid H\in 
\mathcal{H}^{W}\}$. As it happens in the literature for $B=0$ this optimal
solution exists under additional assumptions on the utility function $u$ (or
similar growth conditions on its conjugate), which are introduced in Section %
\ref{secdual}.

\bigskip

Since the most well--studied utility function in the class considered in
this paper is the exponential utility, we specialize the duality result for
this case in Section 3.6, thereby obtaining a generalization of the results
in Bellini and Frittelli \cite{bef}, the "Six Authors paper" \cite{d6} and
Becherer \cite{bec}. Some interesting examples of exponential utility
optimization with random endowment are presented, where the singular part
shows up. These examples are simple, one period market models, but
surprising since they display a quite different behavior from the locally
bounded case, which is thoroughly interpreted.

\bigskip

While the notion of the indifference price was introduced in 1989 by Hodges
and~Neuberger \cite{hn}, the analysis of its dual representation in terms of
(local) martingale measures was performed in the late '90. It started with
Frittelli \cite{f} and was considerably expanded by \cite{d6} and, in a
dynamic context, by El Karoui and Rouge \cite{ekr}. An extensive survey of
the recent literature on this topic can be found in \cite{C08}, Volume on
Indifference Pricing.

Armed with the duality result of Theorem \ref{conj1}, the indifference price
of a claim $B$ is addressed in Section 4. The classical approach of Convex
Analysis - basically the Fenchel-Moreau Theorem - was first applied in
Frittelli and Rosazza \cite{fr} to deduce the dual representation of convex
risk measures on $L^{p}$ spaces. Based on the duality results proven in \ 
\cite{f}, in \cite{fr} it is also shown that, for the exponential utility
function, the indifference price of a bounded claim defines - except for the
sign - a convex risk measure. In recent years this connection has been
deeply investigated by many authors (see Barrieu and N. El Karoui \cite{bnek}
and the references therein). \newline
\indent   In Section 4 of this paper these results are further extended
thanks to the Orlicz space duality framework. This enables us to establish
the properties of the indifference price $\pi$ summarized in Proposition
4.4, including the expected convexity, monotonicity, translation invariance
and volume asymptotics. More interestingly, in \eqref{pi} we provide a new
and fairly explicit representation for the indifference price, which is
obtained applying recent results from the theory of convex risk measures
developed in Biagini and Frittelli \cite{BiaFri07}. In fact, in Proposition
4.4 it is also shown that the map $\pi$, as a convex monotone functional on
the Orlicz space $L^{\widehat{u}}$, is continuous and subdifferential on the
interior of its proper domain $\mathcal{B} $, which is considerably large as
it coincides with $- \mathrm{int}(\mathrm{Dom}(I_u))$, i.e. the opposite of
the interior of the proper domain of the integral functional $I_u(f)
=E[u(f)] $ in $L^{\widehat{u}}$. The minus sign is only due to the fact that 
$\pi(B)$ is the seller indifference price. In Corollary \ref{CorRho} we show
that when $B$ and the loss control $W$ are "very nice" (i.e., they are in
the special subspace $M^{\widehat{u}}$ of $L^{\widehat{u}}$), the
indifference price $\pi$ has also the Fatou property. \newline
\indent  The regularity of the map $\pi$ itself allows then for a very nice,
short proof of some bounds on the indifference price $\pi(B)$ of a fixed
claim $B$ as a consequence of the Max Formula in Convex Analysis.

\bigskip

Section 5 concludes the paper with a comparison with the existing literature
on utility maximization in incomplete semimartingale markets with random
endowment (the reader is deferred to \cite{BiaFri06}\ and the literature
therein for the case of no random endowment) and utility functions finite
valued on $\mathbb{R}$ (see Hugonnier and Kramkov \cite{huk} and the
literature therein for utility functions finite valued on $\mathbb{R}_{+}).$

\section{The set up for utility maximization}

In this section we recall the set up of \cite{BiaFri06} for the utility
maximization problem in an Orlicz space framework with zero random
endowment, corresponding to the left--hand side of \eqref{def_indiff}.
Similar arguments can then be used in the next section for the optimization
problem in the presence of a random endowment as in \eqref{1}. In
particular, the class of admissible integrands as well as the relevant
Orlicz spaces and dual variables are the same for both problems.

\subsection{Admissible integrands, suitability and compatibility}

Given a non--negative random variable $W\in \mathcal{F}_{T}$, the domain of
optimization for the primal problem (\ref{1}) is the following set of $W$--%
\emph{admissible strategies}: 
\begin{equation}
\mathcal{H}^{W}:=\left\{ H\in L(S)\mid \exists c>0\text{\mbox{ such that }}%
\int_{0}^{t}H_{s}dS_{s}\geq -cW,\forall t\in \lbrack 0,T]\right\} ,
\label{HW}
\end{equation}%
where $L(S)$ denotes the class of predictable, $S$-integrable processes. In
other words, the random variable $W$ controls the losses in trading. This
extension of the classic notion of admissibility, which requires $W=1$, was
already used in Schachermayer (\cite{s94} Section 4.1) in the context of the
fundamental theorem of asset pricing, as well as in Delbaen and
Schachermayer \cite{ds6}. \newline
\indent In order to build a reasonable utility maximization, $W$ should
satisfy two conditions that are mathematically useful and economically
meaningful. The first condition depends only on the vector process of traded
assets $S$ and guarantees that the set of $W$--admissible strategies is rich
enough for trading purposes:

\begin{definition}
\label{viability}We say that a random variable $W\geq 1$ is \emph{suitable}
for the process $S$ if for each $i=1,\ldots ,d$, there exists a process $%
H^{i}\in L(S^{i})$ such that 
\begin{equation}
P(\{\omega \mid \exists t\geq 0\ \text{\mbox{such that }}H_{t}^{i}(\omega
)=0\})=0  \label{suit1}
\end{equation}%
and 
\begin{equation}
\left\vert \int_{0}^{t}H_s^{i}dS_s^{i} \right\vert \leq W,\quad \forall t\in
\lbrack 0,T].  \label{suit2}
\end{equation}%
The class of suitable random variables is denoted by $\mathbb{S}.$
\end{definition}

The second condition depends only on the utility function and measures to
what extent the investor accepts the risk of a large loss:

\begin{definition}
\label{compatibility}We say that a positive random variable $W$ is \emph{\
strongly compatible} with the utility function $u$ if 
\begin{equation}
E[u(-\alpha W)]>-\infty \mbox{ for all }\alpha >0  \label{compatible}
\end{equation}%
and that it is \emph{\ compatible} with $u$ if 
\begin{equation}
E[u(-\alpha W)]>-\infty \mbox{ for some }\alpha >0.  \label{w_compatible}
\end{equation}
\end{definition}

Given a suitable and compatible random variable $W$, the first step to apply
duality arguments to problem (\ref{1}) is to rewrite it in terms of an
optimization over random variables, as opposed to an optimization over
stochastic processes. To this end, we define the set of terminal values
obtained from $W$--admissible trading strategies as 
\begin{equation}
K^{W}=\left\{ \int_0^T H_t dS_t\mid H\in \mathcal{H}^{W}\right\} ,
\label{k_def}
\end{equation}%
and consider the modified primal problem 
\begin{equation}
\sup_{k\in K^{W}}E[u(x+k)].  \label{primal_rv}
\end{equation}

The next step is to identify a good dual system and invoke some duality
principle. Classically, the system $(L^{\infty },ba)$ has been successfully
used when dealing with locally bounded traded assets. In order to
accommodate more general markets and inspired by the compatibility
conditions above, in the next section we argue instead for the use of an
appropriate Orlicz spaces duality, naturally induced by the utility function.

\begin{remark}
When $S$ is locally bounded, $W=1$ is automatically suitable and compatible
(see \cite{biafri05}, Proposition 1), and we recover the familiar set of
trading strategies. Therefore, the locally bounded setup is a special case
of our more general framework.
\end{remark}

\begin{remark}
The conditions of suitability and compatibility on $W$ put integrability
restrictions on the jumps of the semimartingale $S$. For a toy example that
illustrates the various situations, see \cite[Example 4]{BiaFri06}.
\end{remark}

\begin{remark}
It is not difficult (see for instance Biagini \cite{b04}, where the utility
maximization for possibly non locally bounded semimartingales was addressed
with a new class of strategies) to build a different set up, where the
definitions of admissibility, suitability and compatibility are formulated
in terms of stochastic processes, instead of random variables, leading to an 
\emph{adapted} control of the losses from trading. \newline
\indent A real, adapted and nonnegative process $Y$ could be defined to be
suitable to $S$ if for each $i=1,\ldots ,d$, there exists a process $%
H^{i}\in L(S^{i})$ satisfying (\ref{suit1}) and $|H^{i}dS^{i}|\leq Y$,
and to be compatible with $u$ if 
\begin{equation*}
E[u(-\alpha Y_{T}^{\ast })]>-\infty \mbox{ for some }\alpha >0,
\end{equation*}%
where $Y_{t}^{\ast }=\sup_{s\leq t}|Y_{s}|$ is the maximal process of $Y$.
The admissible integrands become then 
\begin{equation*}
\mathcal{H}^{Y}:=\{H\in L(S)\mid \exists c>0\text{\mbox{ such that
}}HdS\geq -cY\}.
\end{equation*}%
It is then easy to check that if a process $Y$ satisfies the two
requirements above, then the random variable $W:=Y_{T}^{\ast }$ is suitable
and compatible, in the sense of the Definitions \ref{viability}, \ref%
{compatibility}, and that $\mathcal{H}^{Y}\subset \mathcal{H}^{W}.$ This
shows that this set up with processes does not achieve more generality than
that one with random variables and for this reason we continue to use the
framework described in (\ref{HW}) and in Definitions \ref{viability} and \ref%
{compatibility}.
\end{remark}

\begin{remark}
Alternatively, the same definition of suitability as in the previous remark
could be used, but the process $Y$ could be defined to be compatible with $u 
$ if it satisfies the following less stringent condition: 
\begin{equation}
E[u(-\alpha _{t}Y_{t})]>-\infty \text{\mbox{ for some
}}\alpha _{t}>0,\text{ for all }t\in \lbrack 0,T].  \label{ww}
\end{equation}%
The problem with this definition is that in general (\ref{ww}) does not
guarantee the existence of a uniform bound (in the form of a single random
variable) on the stochastic integrals $\int H_t dS_t $ satisfying the
integrability condition required for the Ansel and Stricker Lemma \cite{as}.
To the best of our knowledge, without this latter result one cannot show
that the regular elements of the dual variables are sigma martingale
measures, a key property that justifies the interpretation of the dual
variables as pricing measures (see the subsequent Section \ref{SecDualV} or 
\cite{ds5}, \cite[Prop. 6]{b04}, \cite[Prop. 6]{biafri05} and \cite[Prop. 19]%
{BiaFri06}).
\end{remark}

\subsection{The Orlicz space framework}

This new framework for utility maximization was first introduced by Biagini 
\cite{biasc} and then considerably expanded in \cite{BiaFri06}, upon which
this section is mostly based. The key observation is that the function $%
\widehat{u}:\mathbb{R}\rightarrow \lbrack 0,+\infty )$ defined as 
\begin{equation*}
\widehat{u}(x)=-u(-|x|)+u(0),
\end{equation*}%
is a Young function (a reference book is \cite{RR}). Thus, its corresponding
Orlicz space 
\begin{equation*}
L^{\widehat{u}}(\Omega ,\mathcal{F},P)=\{f\in L^{0}(\Omega ,\mathcal{F}%
,P)\mid E[\widehat{u}(\alpha f)]<+\infty \mbox{ for
some }\alpha >0\},
\end{equation*}%
is a Banach space (and a Banach lattice) when equipped with the Luxemburg
norm 
\begin{equation}
N_{\widehat{u}}(f)=\inf \left\{ c>0\mid E\left[ \widehat{u}\left( \frac{f}{c}%
\right) \right] \leq 1\right\} .  \label{lux}
\end{equation}%
Since the probability space $(\Omega ,\mathcal{F},P)$ is fixed throughout
the paper, set $L^{p}:=L^{p}(\Omega ,\mathcal{F},P)$, $p\in \lbrack
0,+\infty ],$ and $L^{\widehat{u}}:=L^{\widehat{u}}(\Omega ,\mathcal{F},P).$
Under our assumptions on the utility $u$, it is not difficult to see that $%
L^{\infty }\subseteq {L}^{\widehat{u}}\subseteq L^{1}$. Next consider the
subspace of "very integrable" elements in $L^{\widehat{u}}$ 
\begin{equation*}
M^{\widehat{u}}:=\{f\in L^{\widehat{u}}\mid E[\widehat{u}(\alpha f)]<+\infty %
\mbox{ for all }\alpha >0\}.
\end{equation*}%
Due to the fact that $\widehat{u}$ is continuous and finite on $\mathbb{R}$, 
$M^{\widehat{u}}$ contains $L^{\infty }$ and moreover it coincides with the
closure of $L^{\infty }$ with respect to the Luxemburg norm. However, the
inclusion $M^{\widehat{u}}\subset L^{\widehat{u}}$ is in general strict,
since bounded random variables are not necessarily dense in $L^{\widehat{u}}$
(see \cite[Prop. III.4.3 and Cor. III.4.4]{RR}). This will play a central
role in our work.

As observed in \cite{biasc} and \cite{BiaFri06}, the Young function $%
\widehat{u}$ carries information about the utility on large losses, in the
sense that for $\alpha >0$ we have 
\begin{equation}
E[\widehat{u}(\alpha f)]<+\infty \qquad \Longleftrightarrow \qquad
E[u(-\alpha |f|)]>-\infty ,  \label{uu}
\end{equation}%
a characterization that will be repeatedly used in what follows. For
instance, using \eqref{uu} it is easy to see that

\begin{itemize}
\item \emph{a positive random variable $W$ is strongly compatible (resp.
compatible) with the utility function $u$ if and only if $W\in M^{\widehat{u}%
}$ (resp. $W\in L^{\widehat{u}}$).}
\end{itemize}

When $W\in L^{\widehat{u}}$, the negative part of each element in $K^{W}$
belongs to $L^{\widehat{u}}$, but in general we do not have the inclusion $%
K^{W}\subseteq L^{\widehat{u}}$.

\bigskip

>From the definition of $\Phi $ we know that $\Phi (0)=u(+\infty ),$ $\Phi $
is bounded from below and it satisfies $\lim_{y\rightarrow +\infty }\frac{%
\Phi (y)}{y}=+\infty .$ This limit is a consequence of $u$ being finite
valued on $\mathbb{R}$. Indeed, from the inequality $\Phi (y)\geq u(x)-xy$
for all $x,y\in \mathbb{R}$, we get $\lim \inf_{y\rightarrow +\infty }\frac{%
\Phi (y)}{y}\geq \lim \inf_{y\rightarrow +\infty }\frac{u(x)}{y}-x=-x$ for
all $x\in \mathbb{R}.$

The convex conjugate of $\widehat{u}$, called the \emph{complementary} Young
function in the theory of Orlicz spaces, is denoted here by $\widehat{\Phi }$%
, since it admits the representation 
\begin{equation}
\widehat{\Phi }(y)=\left\{ 
\begin{array}{cc}
0 & \text{ if }|y|\leq \beta \\ 
\Phi (|y|)-\Phi (\beta ) & \text{ if }|y|>\beta%
\end{array}%
\right.  \label{phiphi}
\end{equation}%
where $\beta \geq 0$ is the right derivative of $\widehat{u}$ at $0$, namely 
$\beta =D^{+}\widehat{u}(0)=D^{-}u(0)$, and $\Phi (\beta )=u(0)$. If $u$ is
differentiable, note that $\beta =u^{\prime }(0)$ and it is the unique
solution of the equation $\Phi^{\prime }(y)=0$.

>From (\ref{phiphi}) it then follows that $\widehat{\Phi }$ is also a Young
function, which induces the Orlicz space $L^{\widehat{\Phi }}$ endowed with
the Orlicz (dual) norm 
\begin{equation*}
\Vert g\Vert _{\widehat{\Phi }}=\sup \{E[|fg|]\mid E[\widehat{u}(g)]\leq 1\}.
\end{equation*}%
As before, $L^{\infty }\subseteq {L}^{\widehat{\Phi}}\subseteq L^{1}$.
Moreover, $L^{\Phi}$ is a dual space, as 
\begin{equation}
(M^{\widehat{u}})^{\ast }=L^{\widehat{\Phi }},  \label{Mdual}
\end{equation}%
in the sense that if $Q\in (M^{\widehat{u}})^{\ast }$ is a continuous linear
functional on $M^{\widehat{u}}$, then there exists a unique $g\in L^{%
\widehat{\Phi }}$ such that 
\begin{equation*}
Q(f)=\int_{\Omega }fgdP,\qquad f\in M^{\widehat{u}},
\end{equation*}%
with 
\begin{equation*}
\Vert Q\Vert _{(M^{\widehat{u}})^{\ast }}:={\sup_{N_{\widehat{u}}(f)\leq
1}|Q(f)|}=\Vert g\Vert _{\widehat{\Phi }}.
\end{equation*}%
The characterization of the topological dual for the larger space $L^{%
\widehat{u}}$ is more demanding than (\ref{Mdual}). For the complementary
pair of Young functions $(\widehat{u},\widehat{\Phi })$, it follows from 
\cite[Cor. IV.2.9]{RR} that each element $Q\in (L^{\widehat{u}})^{\ast }$
can be uniquely expressed as 
\begin{equation*}
Q=Q^{r}+Q^{s},
\end{equation*}%
where the \emph{regular} part $Q^{r}$ is given by 
\begin{equation*}
Q^{r}(f)=\int_{\Omega }fgdP,\qquad f\in L^{\widehat{u}},
\end{equation*}%
for a unique $g\in L^{\widehat{\Phi }}$, and the \emph{singular} part $Q^{s}$
satisfies 
\begin{equation}
Q^{s}(f)=0,\qquad \forall f\in M^{\widehat{u}}.  \label{M0}
\end{equation}%
In other words, 
\begin{equation*}
(L^{\widehat{u}})^{\ast }=(M^{\widehat{u}})^{\ast }\oplus (M^{\widehat{u}%
})^{\perp }
\end{equation*}%
where $(M^{\widehat{u}})^{\perp }=\{z\in (L^{\widehat{u}})^{\ast }\mid
z(f)=0,\forall f\in M^{\widehat{u}}\}$ denotes the annihilator of $M^{%
\widehat{u}}$.

Consider now the concave integral functional $I_{u}:L^{\widehat{u}%
}\rightarrow \lbrack -\infty ,\infty )$ defined as 
\begin{equation*}
I_{u}(f):=E[u(f)].
\end{equation*}%
As usual, its effective domain is denoted by 
\begin{equation*}
\mathrm{Dom}(I_{u}):=\left\{ f\in L^{\widehat{u}}\mid E[u(f)]>-\infty
\right\} .
\end{equation*}%
It was shown in \cite[Lemma 17]{BiaFri06} that thanks to the selection of
the appropriate Young function $\widehat{u}$ associated with the utility
function $u$, the norm of a \emph{nonnegative} singular element $z\in (M^{%
\widehat{u}})^{\perp }$ satisfies 
\begin{equation}
\Vert z\Vert _{(L^{\widehat{u}})^{\ast }}:=\sup_{N_{\widehat{u}}(f)\leq
1}z(f)=\sup_{f\in \mathrm{Dom}(I_{u})}z(-f).  \label{Q}
\end{equation}


\subsection{Loss and dual variables\label{sec23}}

\label{SecDualV}

>From now on, the loss controls $W$ are assumed suitable and compatible, i.e. 
$W\in \mathbb{S}\cap L^{\widehat{u}}$, and will simply be referred to as 
\emph{loss variables}. Given such $W$, the cone 
\begin{equation*}
C^{W}=(K^{W}-L_{+}^{0})\cap L^{\widehat{u}},
\end{equation*}%
corresponds to random variables that can be super--replicated by trading
strategies in $\mathcal{H}^{W}$ and that satisfy the same type of
integrability condition of $W$. The polar cone of $C^{W}$, which will play a
role in the dual problem, is 
\begin{equation}
(C^{W})^{0}:=\left\{ Q\in (L^{\widehat{u}})^{\ast }\mid Q(f)\leq 0,\quad
\forall f\in C^{W}\right\} ,  \label{polar_cone}
\end{equation}%
and it satisfies $(C^{W})^{0}\subseteq (L^{\widehat{u}})_{+}^{\ast }$, since 
$(-L_{+}^{\widehat{u}})\subseteq C^{W}$. Therefore, all the functionals of
interest are positive and the decomposition $Q=Q^{r}+Q^{s}$ enables the
identification of $Q^{r}$ with a measure with density $\frac{dQ^{r}}{dP}\in
L_{+}^{\widehat{\Phi }}\subseteq L_{+}^{1}$. The subset of normalized
functionals in $(C^{W})^{0}$ is defined by 
\begin{equation}
\mathcal{M}^{W}:=\{Q\in (C^{W})^{0}\mid Q(\mathbf{1}_{\Omega })=1\}.
\label{normalized}
\end{equation}%
Using the notation above, we see that this normalization condition reduces
to $Q^{r}(\mathbf{1}_{\Omega })=1$, since $Q^{s}\in (M^{\widehat{u}})^{\bot }
$ and thus vanishes on any bounded random variable. In other words, \textit{%
the regular part of any element in} $\mathcal{M}^{W}$ \textit{is a true
probability measure} \textit{with density in} $L_{+}^{\widehat{\Phi }}$.
Moreover, it was shown in \cite[Proposition 19]{BiaFri06}, that 
\begin{equation}
\mathcal{M}^{W}\cap L^{1}=\mathbb{M}_{\sigma }\cap L^{\widehat{\Phi }},
\label{measures}
\end{equation}%
where%
\begin{equation*}
\mathbb{M}_{\sigma }=\left\{ Q(\mathbf{1}_{\Omega })=1,\frac{dQ}{dP}\in
L_{+}^{1}\mid S\text{ is a }\sigma -\text{martingale w.r.t. }Q\right\} 
\end{equation*}%
consists of all the $P-$absolutely continuous $\sigma $-martingale measures
for $S$, i.e. of those $Q\ll P$ for which there exists a process $\eta \in
L(S)$ such that $\eta >0$ and the stochastic integral $\int \eta dS$ is a $Q$%
--martingale. Such probabilities $Q$ were introduced in the context of
Mathematical Finance by Delbaen and Schachermayer in the seminal \cite{ds5},
which the reader is referred to for a thorough analysis of their financial
significance as pricing measures.\newline
\indent From (\ref{measures}) it follows that the regular elements of the
normalized set $\mathcal{M}^{W}$ coincide with the $\sigma $-martingale
measures for $S$ that belong to $L^{\widehat{\Phi }}$. In particular\emph{\
this shows that the (possibly empty) set $\mathcal{M}^{W}\cap L^{1}$ does
not depend on the particular loss variable $W$.}

\subsection{Utility optimization with no random endowment}

The following theorem is a reformulation of \cite[Theorem 21]{BiaFri06}.
When $S$ is locally bounded, a duality formula similar to (\ref{ee}) - but
with no singular components - holds true for all utility functions in the
class considered in this paper. This latter fact is well known and was first
shown in \cite{bef}.

\begin{theorem}
\label{supmin}Suppose that there exists a loss variable $W$ satisfying 
\begin{equation*}
\sup_{H\in \mathcal{H}^{W}}E\left[ u\left( x+\int_{0}^{T}H_{t}dS_{t}\right) %
\right] <u(+\infty ).
\end{equation*}%
Then $\mathcal{M}^{W}$ is not empty and 
\begin{equation}
\sup_{H\in \mathcal{H}^{W}}E\left[ u\left( x+\int_{0}^{T}H_{t}dS_{t}\right) %
\right] =\min_{\lambda >0,\text{ }Q\in \mathcal{M}^{W}}\left\{ \lambda x+E%
\left[ \Phi \left( \lambda \frac{{dQ}^{r}}{{dP}}\right) \right] +\lambda
\Vert Q^{s}\Vert \right\} .  \label{ee}
\end{equation}%
When $W\in M^{\widehat{u}}$, then the set $\mathcal{M}^{W}$ can be replaced
by $\mathbb{M}_{\sigma }\cap L^{\widehat{\Phi }}$ and no singular term
appears in the duality formula above.
\end{theorem}

The last statement in the theorem follows from the observation that when $%
W\in M^{\widehat{u}}$ then the regular component $Q^{r}$ of $Q\in \mathcal{M}%
^{W}$ is already in $\mathcal{M}^{W}$ (see \cite[Lemma 41]{BiaFri06}). Since 
$\Vert Q^{s}\Vert \geq 0$ this immediately implies that the minimum in (\ref%
{ee}) is reached on the set $\left\{ Q^{r}\mid Q\in \mathcal{M}^{W}\right\} =%
\mathbb{M}_{\sigma }\cap L^{\widehat{\Phi }}$.

\section{Utility optimization with random endowment}

\label{claim_section}

\subsection{Conditions on the claim\label{secClaim}}

We now turn to the right--hand side of \eqref{def_indiff} and consider the
optimization problem 
\begin{equation}
\sup_{H\in \mathcal{H}^{W}}E\left[ u\left(
x+\int_{0}^{T}H_{t}dS_{t}-B\right) \right] ,  \label{random}
\end{equation}%
where $B\in \mathcal{F}_{T}$ is a liability faced at terminal $T$.\newline

\indent Without loss of generality, let $x=0$. The case with non null
initial endowment can clearly be recovered by replacing $B$ with $(B-x)$. In
view of the substitution of terminal wealths $\int_{0}^{T}H_tdS_t\in K^{W}$
by random variables $f\in C^{W}\subset L^{\widehat{u}}$, we require that $B$
satisfies 
\begin{equation}
E[u(f-B)]<+\infty ,\qquad \forall f\in L^{\widehat{u}},  \label{gains}
\end{equation}%
so that the concave functional $I_{u}^{B}:L^{\widehat{u}}\rightarrow \lbrack
-\infty ,\infty )$ given by 
\begin{equation*}
I_{u}^{B}(f):=E[u(f-B)]
\end{equation*}%
is well defined for such claims.

\begin{remark}
The set of claims satisfying this condition is quite large. In fact, by
monotonicity and concavity of $u$, \label{negative_stuff} 
\begin{equation*}
E[u(f-B)]=E[u(f-(B^{+}-B^{-}))]\leq E[u(f+B^{-})]\leq u(E[f]+E[B^{-}]),
\end{equation*}%
where the last step follows from Jensen's inequality. Therefore, since $f\in
L^{\hat{u}}\subset L^{1}$, one obtains that a simple sufficient condition
for (\ref{gains}) is that $B^{-}\in L^{1}$.

Obviously, when the utility function is bounded above (as for example in the
exponential case) the condition \eqref{gains} is satisfied by \emph{any}
claim.
\end{remark}

A second natural condition on $B$ is that it does not lead to prohibitive
punishments when the agent chooses the trading strategy $H\equiv 0\in 
\mathcal{H}^{W}$. In other words, we would like to impose that $%
E[u(-B)]>-\infty $. Since the utility function is finite and increasing,
this is equivalent to 
\begin{equation}
E[u(-B^{+})]>-\infty ,  \label{positive_bound}
\end{equation}%
which in turn implies that $-B^{+}\in \mathrm{Dom}(I_{u})$ and consequently $%
B^{+}\in L^{\hat{u}}$, in view of \eqref{uu}. Be aware that $B^{+}\in L^{%
\hat{u}}$ does not necessarily imply (\ref{positive_bound}).

\bigskip

However, for the main duality result we also need that the claim $B$
satisfies: 
\begin{equation}
E[u(-(1+\epsilon )B^{+})]>-\infty ,\qquad \text{ for some }\epsilon >0.
\label{e}
\end{equation}%
This condition, stronger than (\ref{positive_bound}), is equivalent to
requiring that the random variable $(-B^{+})$ belongs to $\mathrm{int}(%
\mathrm{Dom}(I_{u})),$ the interior in $L^{\widehat{u}}$ of the effective
domain of $I_{u}.$ This is a consequence of Lemma 30 in \cite{BiaFri06},
which in turn is based on the definition of the Luxemburg norm on $L^{%
\widehat{u}}$ and on a simple convexity argument. In addition to its
technical relevance (shown in Lemma \ref{continuity}), another reason for
adopting (\ref{e}) is explained in Remark \ref{remB}.

\begin{definition}
\label{defA}The set of \emph{admissible} claims $\mathcal{A}_{u}$ consists
of $\mathcal{F}_{T}$ measurable random variables $B$ satisfying (\ref{gains}%
) and (\ref{e}).
\end{definition}

The conditions (\ref{gains}), \eqref{e} do not really capture the risks
corresponding to $B^{-}$, which are \emph{gains} for the seller of the
claim. For example, it is quite possible to have $B\in \mathcal{A}_{u}$ and $%
E[u(-\varepsilon B^{-})]=-\infty $ for all $\varepsilon >0$ (simply take $%
B^{-}\in L^{1}\backslash L^{\hat{u}}$). This would mean that a \emph{buyer}
using the same utility function $u$, investment opportunities $S$ and
control random variable $W$, would incur losses leading to an infinitely
negative expected utility simply by holding any fraction of the claim and
doing no other investment. Such undesirable outcome can be avoided by the
more stringent condition 
\begin{equation}
E[u(-\varepsilon B^{-})]>-\infty ,\qquad \mbox{ for some }\varepsilon >0,
\label{negative_bound}
\end{equation}%
which is equivalent to $B^{-}\in L^{\hat{u}}$. Since the focus is on the
problem faced by the seller of the claim $B$, \textbf{we refrain from
assuming \eqref{negative_bound}, until Section \ref{sec4} where $B$ will
belong to the set} 
\begin{equation}
\mathcal{B}:=\mathcal{A}_{u}\cap L^{\widehat{u}}=\{B\in L^{\widehat{u}}\mid
E[u(-(1+\epsilon )B^{+})]>-\infty \text{ for some }\epsilon >0\}.  \label{bb}
\end{equation}
In any event, the potential buyers for $B$ will likely not have the same
investment opportunities, utility function and loss tolerance as the seller,
leading to entirely different versions of \eqref{def_indiff}.

For example, suppose that the seller has an exponential utility $%
u_{s}(x)=-e^{-x}$ and the buyer has quadratic utility $u_{b}(x)=-x^{2}$ for $%
x\leq -1$, then prolonged so that it is bounded above and satisfies all the
other requirements. Then take $B$ so that $B^{+}$ has an exponential
distribution with parameter $\lambda =2$ and $B^{-}$ has a density $\frac{c}{%
1+x^{4}},x\geq 0$, where $c$ is the normalizing constant. It is easy to
check that $B $ satisfies \eqref{gains} and \eqref{e} for $u_{s}$ and that
selling $B$ is very attractive, since the tail of the distribution of $B^{-}$
(the gain for the seller) is much bigger than that of $B^{+}$ (the loss for
the seller). $B^{-}$ has no finite exponential moment, therefore it violates %
\eqref{negative_bound} and would clearly be unacceptable if the buyer had
exponential preferences. However the quadratic tolerance of the losses of $%
u_{b}$ accounts for a well posed maximization problem with $B$ even for the
buyer.

\subsection{The maximization}

The first step in our program consists in showing that optimizing over the
cone $C^W$ leads to the same expected utility as optimizing over the set of
terminal wealths $K^W$.

\begin{lemma}
If $B$ satisfies (\ref{gains}) and (\ref{positive_bound})\ then 
\begin{equation}
\sup_{k\in K^{W}}E[u(k-B)]=\sup_{f\in C^{W}}E[u(f-B)].  \label{step1}
\end{equation}
\end{lemma}

\begin{proof}
Since $C^{W}\subset (K^{W}-L_{+}^{0})$ and the utility function is monotone
increasing, 
\begin{equation}
\sup_{f\in C^{W}}E[u(f-B)]\leq \sup_{g\in (K^{W}-L_{+}^{0})}E[u(g-B)]\leq
\sup_{k\in K^{W}}E[u(k-B)].  \label{one_side}
\end{equation}%
Since $k\equiv 0\in K^{W}$, 
\begin{equation*}
\sup_{k\in K^{W}}E[u(k-B)]\geq E[u(-B)]>-\infty .
\end{equation*}%
by (\ref{positive_bound}). Pick any $k\in K^{W}$ satisfying $%
E[u(k-B)]>-\infty $. Consider $k_{n}=k\wedge n$, which is in $C^{W}$ since $%
W\in L^{\widehat{u}}$ (this is the only assumption needed on $W$ here). Then 
\begin{equation*}
u(k_{n}-B)=u(k^{+}\wedge n-B)I_{\{k\geq 0\}}+u(-k^{-}-B)I_{\{k<0\}}\geq
u(-B)+u(-k^{-}-B)I_{\{k<0\}}
\end{equation*}%
and the latter is integrable. An application of the monotone convergence
theorem gives $E[u(k_{n}-B)]\nearrow E[u(k-B)]$, which implies that 
\begin{equation*}
\sup_{k\in K^{W}}E[u(k-B)]\leq \sup_{f\in C^{W}}E[u(f-B)],
\end{equation*}%
and completes the proof.
\end{proof}

\bigskip

The next step in the program is to establish that the functional $I_{u}^{B}$
has a norm continuity point contained in the cone of interest $C^{W}$.

\begin{lemma}
\label{continuity}Suppose that $B$ satisfies (\ref{gains}) and (\ref%
{positive_bound}). Then the concave functional $I_{u}^{B}$ is norm
continuous on the interior of its effective domain. Moreover, if $B\in 
\mathcal{A}_{u}$ then there exists a norm continuity point of $I_{u}^{B}$
that belongs to $C^{W}$.
\end{lemma}

\begin{proof}
Since $I_{u}^{B}<+\infty ,$ the functional $I_{u}^{B}$ is proper, monotone
and concave. The first sentence in the Lemma thus follows from the Extended
Namioka-Klee Theorem (see \cite{ruz} or \cite{BiaFri07}). Denoting the unit
ball in $L^{\hat{u}}$ by $\mathcal{S}_{1}$, it follows rather easily from a
convexity argument that the hypothesis (\ref{e}) on $B^{+}$ implies that 
\begin{equation}
-B^{+}+\frac{\epsilon }{1+\epsilon }\mathcal{S}_{1}\subset \mathrm{Dom}%
(I_{u}),  \label{99}
\end{equation}%
and therefore $\frac{\epsilon }{1+\epsilon }\mathcal{S}_{1}\subset \mathrm{%
Dom}(I_{u}^{B})$. Therefore, any element of $\frac{\epsilon }{2(1+\epsilon )}%
\mathcal{S}_{1}\cap (-L_{+}^{\widehat{u}})$ is then in $C^{W}$ and a
continuity point for $I_{u}^{B}$.
\end{proof}

\begin{remark}
\label{remB}At first sight, the condition \eqref{e} on the positive part $%
B^{+}$ appears to be an ad-hoc hypothesis imposed for the sake of proving
the previous technical lemma. We argue, however, that \eqref{e} is in fact a
natural condition to impose on a financial liability in this context.
Indeed, if the claim $B$ satisfies only $E[u(-B^{+})]>-\infty $, it may
happen - contrary to (\ref{99}) - that 
\begin{equation*}
E[u(-B^{+}-c)]=-\infty \mathrm{\ for\ all\ constants\ }c>0
\end{equation*}%
as shown in the example below, which would restrict the possibility of any
significative trading.
\end{remark}

\begin{example}
Consider the smooth function 
\begin{equation*}
u(x)=\left\{ 
\begin{array}{cc}
-e^{(x-1)^{2}} & x\leq 0 \\ 
-2e^{-x+1}+e & x>0%
\end{array}%
\right.
\end{equation*}%
as utility function $u$ (the particular expression of $u$ for $x>0$ is
however irrelevant). Consider now a (positive) claim $B$ with distribution $%
d\mu _{B}=k\frac{e^{-x^{2}-2x}}{x^{2}+1}I_{\{x>0\}}dx$, where $k$ is the
normalizing constant. Then, 
\begin{equation*}
E[u(-B)]=\int_{0}^{+\infty }-e^{(-x-1)^{2}}k\frac{e^{-x^{2}-2x}}{x^{2}+1}%
dx>-\infty
\end{equation*}%
but for any $c>0$, 
\begin{equation*}
E[u(-B-c)]=\int_{0}^{+\infty }-\frac{e^{2cx+(c+1)^{2}}k}{x^{2}+1}dx=-\infty .
\end{equation*}
\end{example}

\subsection{Conjugate functionals}

As already discussed, the condition $B\in \mathcal{A}_{u}$ on the claim $B$
does not necessarily imply that $B\in L^{\widehat{u}}$. Therefore, we need
appropriate extensions of linear functionals on $L^{\widehat{u}}$.

Though we sketch the proof for the sake of completeness, this extension is
morally straightforward. In fact, it is defined in the same way as the
expectation $E[g]$ is defined when $g$ is bounded from below, instead of
bounded. In this case, $g^{-}\in L^{\infty }$ and 
\begin{equation*}
E[g]:=\sup \{E[f]\mid f\in L^{\infty },f\leq g\}=\lim_{n}E[g\wedge n]
\end{equation*}%
Accordingly, let us consider the convex cone of random variables with
negative part in $L^{\widehat{u}}$: 
\begin{equation*}
L_{neg}^{\widehat{u}}:=\left\{ f\in L^{0}\mid f^{-}\in L^{\widehat{u}%
}\right\} =\left\{ f\in L^{0}\mid E[u(-\alpha f^{-})]>-\infty ,%
\mbox{ for
some }\alpha >0\right\} ,
\end{equation*}%
and notice that this cone contains $K^{W},$ for any loss variable $W$. For
any $Q\in (L^{\widehat{u}})_{+}^{\ast }$ we define $\widehat{Q}:L_{neg}^{%
\widehat{u}}\rightarrow \mathbb{R}\cup \left\{ +\infty \right\} $ by 
\begin{equation}
\widehat{Q}(g)\triangleq \sup \left\{ Q(f)\mid f\in L^{\widehat{u}}\text{
and }f\leq g\right\} .  \label{ZZ}
\end{equation}

\begin{lemma}
\label{lemmaHat}If $Q\in (L^{\widehat{u}})_{+}^{\ast }$ then

\begin{enumerate}
\item $\widehat{Q}$ is a well-defined extension of $Q$. It is a positively
homogenous, additive (with the convention $+\infty +c=+\infty $ for $c\in 
\mathbb{R}\cup \{+\infty \}$), monotone functional on the cone $L_{neg}^{%
\widehat{u}}$. In particular, if $g\in L_{neg}^{\widehat{u}},h\in L^{%
\widehat{u}}$ then $\widehat{Q}(g+h)=\widehat{Q}(g)+Q(h)$.

\item $Q\in (C^{W})^{0}$ if and only if $\widehat{Q}(k)\leq 0$ for all $k\in
K^{W}$.

\item If $g\in L_{neg}^{\widehat{u}}$ is such that $E[u(g)]>-\infty $, then 
\begin{equation}
\Vert Q^{s}\Vert \geq -\widehat{Q^{s}}(g)  \label{singineq}
\end{equation}

\item If $\widehat{Q}(g)$ is finite, then $\widehat{Q}(g)=E[\frac{dQ^{r}}{dP}%
g]+ \widehat{Q^{s}}(g)$.
\end{enumerate}
\end{lemma}

\begin{proof}
The first two statements and item 4 follow rather directly from the
definitions of $\widehat{Q}$ and $C^{W}$. We only prove additivity of $%
\widehat{Q}$. Fix then $g_{1},g_{2}\in L_{neg}^{\widehat{u}}$. We want to
show that $\widehat{Q}(g_{1}+g_{2})=\widehat{Q}(g_{1})+\widehat{Q}(g_{2}).$
When $f_{i}\in L^{\widehat{u}}$, $i=1,2,$ satisfy $f_{i}\leq g_{i}$, 
\begin{equation*}
\widehat{Q}(g_{1})+\widehat{Q}(g_{2})=\sup_{f_{i}\leq g_{i}}\left\{
Q(f_{1})+Q(f_{2})\right\} \leq \sup_{f\leq g_{1}+g_{2}}Q(f)=\widehat{Q}%
(g_{1}+g_{2}).
\end{equation*}%
To show the opposite inequality, assume first that $g_{i}\geq 0$. Fix $f\in
L_{+}^{\widehat{u}}$, $f\leq g_{1}+g_{2}$. Then $f\wedge g_{i}\in L^{%
\widehat{u}}$, $i=1,2,$ and, moreover, $f\leq f\wedge g_{1}+f\wedge g_{2}$.
Therefore 
\begin{equation*}
Q(f)\leq Q(f\wedge g_{1})+Q(f\wedge g_{2})\leq \widehat{Q}(g_{1})+\widehat{Q}%
(g_{2})\text{ for all }f\in L_{+}^{\widehat{u}},\text{ }f\leq g_{1}+g_{2}
\end{equation*}%
so that $\widehat{Q}(g_{1}+g_{2})\leq \widehat{Q}(g_{1})+\widehat{Q}(g_{2})$%
. To treat the case $g_{1}$ and $g_{2}$ not necessarily positive, observe
that when $g\in L_{neg}^{\widehat{u}},h\in L^{\widehat{u}}$ :%
\begin{eqnarray*}
\widehat{Q}(g+h) &=&\sup \left\{ Q(f)\mid f\in L^{\widehat{u}}\text{, }f\leq
g+h\right\} \\
&=&\sup \left\{ Q(f)\mid f\in L^{\widehat{u}}\text{, }f\leq g\right\} +Q(h)=%
\widehat{Q}(g)+Q(h),
\end{eqnarray*}%
As a consequence, $\widehat{Q}(g_{i})=\widehat{Q}(g_{i}^{+})-Q(g_{i}^{-})$
and $\widehat{Q}(g_{1}+g_{2})=\widehat{Q}%
(g_{1}^{+}+g_{2}^{+})-Q(g_{1}^{-}+g_{2}^{-})$. Collecting these relations,%
\begin{equation*}
\widehat{Q}(g_{1}+g_{2})=\widehat{Q}%
(g_{1}^{+}+g_{2}^{+})-Q(g_{1}^{-}+g_{2}^{-})\leq \widehat{Q}(g_{1}^{+})+%
\widehat{Q}(g_{2}^{+})-Q(g_{1}^{-}+g_{2}^{-})=\widehat{Q}(g_{1})+\widehat{Q}%
(g_{2}).
\end{equation*}%
\noindent Item 3 follows from $-g^{-}\in \mathrm{Dom}(I_{u})$ and equation %
\eqref{Q}: 
\begin{equation*}
\Vert Q^{s}\Vert \geq Q^{s}(g^{-})\geq Q^{s}(g^{-})-\widehat{Q^{s}}(g^{+})=-%
\widehat{Q^{s}}(g).
\end{equation*}
\end{proof}

Finally, when $B$ satisfies (\ref{positive_bound}) the \emph{convex conjugate%
} $J_{u}^{B}:(L^{\widehat{u}})^{\ast }\rightarrow \mathbb{R}\cup \{+\infty
\} $ of the concave functional $I_{u}^{B}$ is defined as 
\begin{equation}
J_{u}^{B}(Q):=\sup_{f\in L^{\widehat{u}}}\left\{ E[u(f-B)]-Q(f)\right\}
,\quad Q\in (L^{\widehat{u}})^{\ast }.  \label{J}
\end{equation}%
The following Lemma gives a representation of $J_{u}^{B}$.

\begin{lemma}
\label{LemmaJJ}

\begin{enumerate}
\item If $B\in L^{\widehat{u}}$ and $Q\in (L^{\widehat{u}})_{+}^{\ast }$
then 
\begin{equation}
J_{u}^{B}(Q)=Q(-B)+E\left[ \Phi \left( \frac{dQ^{r}}{dP}\right) \right]
+\Vert Q^{s}\Vert .  \label{JJ}
\end{equation}

\item If $B$ satisfies (\ref{gains}) and (\ref{positive_bound})\ and $Q\in
(L^{\widehat{u}})_{+}^{\ast }$, then 
\begin{equation*}
J_{u}^{B}(Q)=\widehat{Q}(-B)+E\left[ \Phi \left( \frac{dQ^{r}}{dP}\right) %
\right] +\Vert Q^{s}\Vert .
\end{equation*}
\end{enumerate}
\end{lemma}

\begin{proof}
\begin{enumerate}
\item This is an elementary consequence of the representation result proved
in \cite[Theorem 2.6]{koz}. In fact, from $B\in L^{\widehat{u}}$, 
\begin{eqnarray*}
J_{u}^{B}(Q) &=&\sup_{f\in L^{\widehat{u}}}\left\{ E[u(f-B)]-Q(f)\right\} \\
&=&\sup_{g\in L^{\widehat{u}}}\left\{ E[u(g)]-Q(g)\right\} -Q(B)
\end{eqnarray*}%
and in the cited Theorem, Kozek proved that 
\begin{equation*}
\sup_{g\in L^{\widehat{u}}}\left\{ E[u(g)]-Q(g)\right\} =E\left[ \Phi \left( 
\frac{dQ^{r}}{dP}\right) \right] +\sup_{g\in \mathrm{Dom}(I_{u})}Q^{s}(-g)
\end{equation*}%
so that the thesis in (\ref{JJ}) is enabled by \eqref{Q}.

\item The thesis follows from the equality 
\begin{eqnarray*}
&&\sup_{G\geq B,G\in L^{\widehat{u}}}\left\{ \sup_{f\in L^{\widehat{u}%
}}\left\{ E[u(f-G)]-Q(f)\right\} \right\} \\
&=&\sup_{f\in L^{\widehat{u}}}\left\{ \sup_{G\geq B,G\in L^{\widehat{u}%
}}\{E[u(f-G)]-Q(f)\}\right\} .
\end{eqnarray*}

Indeed, thanks to \eqref{JJ}, the left hand side gives 
\begin{eqnarray*}
&&\sup_{G\geq B,G\in L^{\widehat{u}}}\left\{ \sup_{f\in L^{\widehat{u}%
}}\left\{ E[u(f-G)]-Q(f)\right\} \right\} \\
&=&\sup_{G\geq B,G\in L^{\widehat{u}}}J_{u}^{G}(Q)=\widehat{Q}(-B)+E\left[
\Phi \left( \frac{dQ^{r}}{dP}\right) \right] +\Vert Q^{s}\Vert ,
\end{eqnarray*}%
while the right hand side gives 
\begin{equation}
\sup_{f\in L^{\widehat{u}}}\left\{ \sup_{G\geq B,G\in L^{\widehat{u}%
}}\{E[u(f-G)]-Q(f)\}\right\} =\sup_{f\in L^{\widehat{u}}}\{E[u(f-B)]-Q(f)%
\}=J_{u}^{B}(Q).  \label{88}
\end{equation}%
The first equality in (\ref{88}) holds thanks to the following approximation
argument. For each $f\in L^{\widehat{u}}$ such that $E[u(f-B)]>-\infty $ let 
$G_{n}:=B^{+}-(f^{-}+n)\wedge B^{-}\in L^{\widehat{u}}$ and $A_{n}:=\left\{
B^{-}\leq f^{-}+n\right\} $. Our assumptions imply that $u(f-B)$ and $%
u(-B^{+})$ are integrable and so from 
\begin{equation*}
u(f-G_{n})=u(f-B)1_{A_{n}}+u(f-B^{+}+f^{-}+n)1_{A_{n}^{c}}\geq
u(f-B)1_{A_{n}}+u(-B^{+})1_{A_{n}^{c}}
\end{equation*}%
we deduce $E[u(f-G_{n})]>-\infty $. Since $-G_{n}\uparrow -B$, the monotone
convergence theorem guarantees $\sup_{n}E[u(f-G_{n})]=E[u(f-B)]$.
\end{enumerate}
\end{proof}

\subsection{The dual optimization and a new primal domain\label{secdual}}

Before establishing the main duality result, let us focus on dual
optimizations of the form 
\begin{equation}
\inf_{\lambda >0,Q\in \mathcal{N}}\left\{ E\left[ \Phi \left( \lambda \frac{%
dQ^{r}}{dP}\right) \right] +\lambda \widehat{Q}(-B)+\lambda \Vert Q^{s}\Vert
\right\} ,  \label{dual_general}
\end{equation}%
where $\mathcal{N}$ is a convex subset of $(L^{\widehat{u}})_{+}^{\ast }$.
Problems of this type (but for $\mathcal{N}\subseteq L_{+}^{1}$ and $B=0$)
were originally solved by Ruschendorf \cite{ru}. A general strategy for
tackling such problems is to consider the minimizations over $\lambda $ and
over $Q$ separately. Accordingly, in the next Proposition we fix $\lambda >0$
and explore the consequences of optimality in $Q$. The result is the
analogue of \cite[Prop 25]{BiaFri06} but in the presence of the claim $B$
and the corresponding extended functionals $\widehat{Q}$.

The presence of the scaling factor $\lambda $ in the expectation term in (%
\ref{dual_general}) leads us to consider the convex set: 
\begin{equation*}
\mathcal{L}_{\Phi }=\{Q\text{ probab, }Q\ll P\mid E\left[ \Phi \left(
\lambda \frac{dQ}{dP}\right) \right] <+\infty \text{ for some }\lambda >0\}.
\end{equation*}%
Clearly $\left\{ \frac{dQ}{dP}\mid Q\in \mathcal{L}_{\Phi }\right\}
\subseteq L_{+}^{\widehat{\Phi }}\subseteq L_{+}^{1}$, but the condition $%
\frac{dQ}{dP}\in L_{+}^{\widehat{\Phi }}$\ does not in general imply $E\left[
\Phi (\frac{dQ}{dP})\right] <+\infty $, nor $Q\in \mathcal{L}_{\Phi }$.
Indeed, the utility function may be unbounded from above, so that $\Phi
(0)=+\infty $ is possible.

\begin{remark}
The fact that the set $\mathcal{L}_{\Phi }$ is convex requires a brief
explanation, since $\Phi (0)=+\infty $ is possible. Let $%
Q_{y}=yQ_{1}+(1-y)Q_{2}$ , $y\in (0,1),$ be the convex combination of any
couple of elements in $\mathcal{L}_{\Phi }$, take $\lambda _{i}>0$
satisfying $E\left[ \Phi \left( \lambda _{i}\frac{dQ_{i}}{dP}\right) \right]
<\infty ,$ $i=1,2,$ and define $z_{y}$ as the convex combination of $\frac{1%
}{\lambda _{1}}$ and $\frac{1}{\lambda _{2}}$, i.e.: $z_{y}:=y\frac{1}{%
\lambda _{1}}+(1-y)\frac{1}{\lambda _{2}}\in (0,\infty )$. As a consequence
of the convexity of the function $(z,k)\rightarrow z\Phi (\frac{1}{z}k)$ on $%
\mathbb{R}_{+}\times \mathbb{R}_{+}$(which has been pointed out by \cite{sw}%
, Section 3) we deduce%
\begin{equation*}
E\left[ z_{y}\Phi \left( \frac{1}{z_{y}}\frac{dQ_{y}}{dP}\right) \right]
\leq y\frac{1}{\lambda _{1}}E\left[ \Phi \left( \lambda _{1}\frac{dQ_{1}}{dP}%
\right) \right] +(1-y)\frac{1}{\lambda _{2}}E\left[ \Phi \left( \lambda _{2}%
\frac{dQ_{2}}{dP}\right) \right] <\infty .
\end{equation*}
\end{remark}

\bigskip

\noindent \textbf{Assumption (A)} \textit{The utility function }$u:\mathbb{R}%
\rightarrow \mathbb{R}$\textit{\ is strictly increasing, strictly concave,
continuously differentiable and it satisfies the conditions} 
\begin{equation}
\lim_{x\downarrow -\infty }u^{\prime }(x)=+\infty \text{, }\lim_{x\uparrow
\infty }u^{\prime }(x)=0  \label{inada}
\end{equation}%
\begin{equation}
\mathcal{L}_{\Phi }=\{Q\text{ probab, }Q\ll P\mid E\left[ \Phi \left(
\lambda \frac{dQ}{dP}\right) \right] <+\infty \text{ for all }\lambda >0\}.
\label{homog}
\end{equation}

The condition expressed in (\ref{homog}) coincides with assumption (3) in 
\cite{BiaFri06}. A detailed discussion on assumption (A) and the
relationship of (\ref{homog}) with the condition of Reasonable Asymptotic
Elasticity introduced by Schachermayer \cite{SCHA01} can be found in \cite%
{BiaFri06}, \cite{biafri05}. We stress that this assumption is needed only
when dealing with the existence of optimal solutions (i.e. only in
Propositions \ref{th111}, \ref{perp}, \ref{hedge}, and in the second part of
Theorem \ref{conj1}).

\begin{proposition}
\label{th111}Suppose that the utility $u$ satisfies assumption $A$ and that
the claim $B$ satisfies (\ref{gains}) and (\ref{positive_bound})\ . Fix $%
\lambda >0$ and suppose that $\mathcal{N}\subseteq (L^{\widehat{u}%
})_{+}^{\ast }$ is a convex set such that for any $Q\in \mathcal{N}$ we have 
$Q^{r}\in \mathcal{L}_{\Phi }$. If $Q_{\lambda }\in \mathcal{N}$ is optimal
for 
\begin{equation}
\inf_{Q\in \mathcal{N}}\left\{ E\left[ \Phi \left( \lambda \frac{dQ^{r}}{dP}%
\right) \right] +\lambda \widehat{Q}(-B)+\lambda \Vert Q^{s}\Vert \right\}
<+\infty  \label{41}
\end{equation}%
then, $\forall Q\in \mathcal{N}$ with $\widehat{Q}(-B)<+\infty $ 
\begin{equation}
E_{Q_{\lambda }^{r}}\left[ \Phi ^{\prime }\left( \lambda \frac{dQ_{\lambda
}^{r}}{dP}\right) \right] +\widehat{Q}_{\lambda }(-B)+\Vert Q_{\lambda
}^{s}\Vert \leq E_{Q^{r}}\left[ \Phi ^{\prime }\left( \lambda \frac{%
dQ_{\lambda }^{r}}{dP}\right) \right] +\widehat{Q}(-B)+\Vert Q^{s}\Vert .
\label{bbb}
\end{equation}
\end{proposition}

\begin{proof}
If $Q_{\lambda }$ is optimal then ${\widehat{Q_{\lambda }}}(-B)$ must be
finite. We can assume $\lambda =1$, the case with general $\lambda $ being
analogous, since condition (\ref{homog}) holds true. Denoting the optimal
functional by $Q_{1}$, fix any $Q$ with $\widehat{Q}(-B)$ finite and
consider $Q_{x}=xQ_{1}+(1-x)Q$. Also denote by $V(Q)$ the objective function
to be minimized in (\ref{41}) when $\lambda =1$. Consider the convex
function of $x$ 
\begin{equation*}
F(x):=E\left[ \Phi \left( \frac{dQ_{x}^{r}}{dP}\right) \right] +x\widehat{%
Q_{1}}(-B)+(1-x)\widehat{Q}(-B)+\Vert Q_{x}^{s}\Vert
\end{equation*}%
then $F(1)=V(Q_{1})$ and $F(x)\geq V(Q_{x})$ since $\widehat{Q_{x}}(-B)$ is
convex in $x$. Taking this inequality into account and given that $Q_{1}$ is
a minimizer of $V(Q)$, 
\begin{equation*}
F^{\prime }(1_{-})\leq V^{\prime }(Q_{1_{-}})\leq 0.
\end{equation*}%
Now, as in \cite[Prop 25]{BiaFri06}, it can be shown that 
\begin{equation*}
F^{\prime }(1_{-})=E\left[ \left( \frac{dQ_{1}^{r}}{dP}-\frac{dQ^{r}}{dP}%
\right) \Phi ^{\prime }\left( \frac{dQ_{1}^{r}}{dP}\right) \right] +\widehat{%
Q_{1}}(-B)-\widehat{Q}(-B)+\Vert Q_{1}^{s}\Vert -\Vert Q^{s}\Vert
\end{equation*}%
and since this quantity must be non positive we conclude the proof.
\end{proof}

\bigskip

Next we fix $Q$ and explore the consequences of optimality in $\lambda $.
The result is identical to \cite[Prop 26]{BiaFri06}, which we reproduce here
for readability:

\begin{proposition}
\label{perp}Suppose that the utility $u$ satisfies assumption $A.$ If $Q$ is
a probability measure in $\mathcal{L}_{\Phi }$ then for all $c\in \mathbb{R}$
the optimal $\lambda (c;Q)$ solution of 
\begin{equation}
\min_{\lambda >0}\left\{ E\left[ \Phi \left( \lambda \frac{dQ}{dP}\right) %
\right] +\lambda c\right\}  \label{minl}
\end{equation}%
is the unique positive solution of the first order condition 
\begin{equation}
E\left[ \frac{dQ}{dP}\Phi ^{\prime }\left( \lambda \frac{dQ}{dP}\right) %
\right] +c=0.  \label{lll}
\end{equation}%
The random variable $f^{\ast }:=-\Phi ^{\prime }(\lambda (c;Q)\frac{dQ}{dP}%
)\in \left\{ f\in L^{1}(Q)\mid E_{Q}[f]=c\right\} $ satisfies $u(f^{\ast
})\in L^{1}(P)$ and 
\begin{eqnarray}
&&\min_{\lambda >0}\left\{ E\left[ \Phi \left( \lambda \frac{dQ}{dP}\right) %
\right] +\lambda c\right\}  \notag \\
&=&\sup \left\{ E[u(f)]\mid f\in L^{1}(Q)\text{ and }E_{Q}[f]\leq c\right\}
=E[u(f^{\ast })]<u(\infty )  \label{minsup}
\end{eqnarray}
\end{proposition}

Therefore, whenever $\widehat{Q}(-B)$ is finite, we can set $c=\widehat{Q}%
(-B)+\Vert Q^{s}\Vert $ and conclude from \eqref{minsup} that the
minimization of the objective function in \eqref{dual_general} with respect
to $\lambda >0$ for a fixed $Q$ leads to the same value of a utility
maximization over integrable functions satisfying $E_{Q}[f]\leq \widehat{Q}%
(-B)+\Vert Q^{s}\Vert $. Motivated by these results, we define the following
set of functionals and corresponding domain for utility maximization:

\begin{definition}
For any $B$ satisfying (\ref{gains}) and (\ref{positive_bound})\ let 
\begin{equation}
\mathcal{N}_{B}^{W}:=\{Q\in \mathcal{M}^{W}\mid Q^{r}\in \mathcal{L}_{\Phi },%
\text{ }\widehat{Q}(-B)\in \mathbb{R}\}  \label{nnn}
\end{equation}%
and 
\begin{equation}
K_{B}^{W}:=\{f\in L^{0}\mid f\in L^{1}(Q^{r}),\text{ }E_{Q^{r}}[f]\leq 
\widehat{Q^{s}}(-B)+\Vert Q^{s}\Vert ,\ \ \forall Q\in \mathcal{N}_{B}^{W}\},
\label{kkk}
\end{equation}%
with the corresponding optimization problem 
\begin{equation}
U_{B}^{W}:=\sup_{f\in K_{B}^{W}}E[u(f-B)].  \label{primal_B}
\end{equation}
\end{definition}

\begin{remark}
\label{NB} (i) Note that the sets $\mathcal{N}_{B}^{W}$ and $K_{B}^{W}$
depend also on $\Phi $ (and thus on the utility $u$), but the dependence is
omitted for convenience of notation. In the particular case $B\in L^{%
\widehat{u}}$ however $\widehat{Q}(-B)=Q(-B)$, so that $\mathcal{N}_{B}^{W}$
does not depend on $B$ and it coincides with the set of dual functionals
used in \cite{BiaFri06}: 
\begin{equation}
\mathcal{N}^{W}=\{Q\in \mathcal{M}^{W}\mid Q^{r}\in \mathcal{L}_{\Phi }\}.
\label{ennew}
\end{equation}%
While we are going to treat the utility maximization with random endowment
for general $B$, we will focus on $B\in L^{\widehat{u}}$ in the indifference
price section, where the set of dual functionals will be simply $\mathcal{N}%
^{W}$. Note also that each element in $\mathcal{N}_{B}^{W}$ has non zero
regular part.

(ii) If $\Phi (0)<+\infty ,$ then $Q\in \mathcal{L}_{\Phi }$ iff $Q$ is a
probability s.t. $\frac{{dQ}}{{dP}}\in L_{+}^{\widehat{\Phi }}$. As
explained in Section \ref{sec23} the regular part of each element in $%
\mathcal{M}^{W}$ is already in $L_{+}^{\widehat{\Phi }}$, so that from (\ref%
{nnn}) we get: 
\begin{equation*}
\mathcal{N}_{B}^{W}:=\{Q\in \mathcal{M}^{W}\mid Q^{r}\neq 0,\text{ }\widehat{%
Q}(-B)\in \mathbb{R}\}.
\end{equation*}

(iii) When Assumption (A) is satisfied then 
\begin{equation*}
\mathcal{N}_{B}^{W}:=\{Q\in \mathcal{M}^{W}\mid \text{ }Q^{r}\neq 0,\text{ }E%
\left[ \Phi \left( \frac{{dQ}^{r}}{{dP}}\right) \right] <+\infty ,\text{ }%
\widehat{Q}(-B)\in \mathbb{R}\}.
\end{equation*}
\end{remark}

The utility optimization over the modified domain $K_B^{W} $ can be easily
related with the original utility optimization over terminal wealths $K^W$.

\begin{lemma}
\label{ineq}Suppose that $B$ satisfies (\ref{gains}) and (\ref%
{positive_bound})\ and that $\mathcal{N}_{B}^{W}\neq \emptyset $. Then $%
K^{W}\subset K_{B}^{W}$ and the following chain of inequalities holds true 
\begin{equation*}
\sup_{k\in K^{W}}E[u(k-B)]\leq U_{B}^{W}\leq \inf_{\lambda >0,Q\in \mathcal{N%
}_{B}^{W}}\left\{ \lambda \widehat{Q}(-B)+E\left[ \Phi \left( \lambda \frac{{%
dQ}^{r}}{{dP}}\right) \right] +\lambda \Vert Q^{s}\Vert \right\} <\infty
\end{equation*}
\end{lemma}

\begin{proof}
For the first inequality, fix a $k\in K^{W}$ such that $E[u(k-B)]>-\infty$ . By Lemma \ref{lemmaHat} item 2, $\widehat{Q%
}(k)\leq 0$ for all $Q\in \mathcal{M}^{W}$. The assumptions on $B$ imply
that $B^{+}\in L^{\widehat{u}}$ so that $(-B)\in L_{neg}^{\widehat{u}}.$ In
addition, $K^{W}\subseteq L_{neg}^{\widehat{u}}$ implies $k-B\in L_{neg}^{%
\widehat{u}}$. Applying Lemma \ref{lemmaHat}, item 1, we have for each $Q\in 
\mathcal{N}_{B}^{W}$ 
\begin{equation*}
\widehat{Q}(k-B)=\widehat{Q}(k)+\widehat{Q}(-B)\leq \widehat{Q}(-B)<+\infty .
\end{equation*}%
Then $\widehat{Q}(k-B)$ is finite and, by Lemma \ref{lemmaHat} item 4, the
above inequality becomes: 
\begin{equation*}
E_{Q^{r}}[k-B]+\widehat{Q^{s}}(k-B)\leq E_{Q^{r}}[-B]+\widehat{Q^{s}}(-B).
\end{equation*}%
Given that $-\widehat{Q^{s}}(k-B)\leq \Vert Q^{s}\Vert $ from %
\eqref{singineq} 
\begin{equation*}
E_{Q^{r}}[k-B]\leq E_{Q^{r}}[-B]+\widehat{Q^{s}}(-B)+\Vert Q^{s}\Vert
\end{equation*}%
and thus, cancelling $E_{Q^{r}}[-B]$, we get $k\in K_{B}^{W}$.

To prove the second inequality, the (pointwise) Fenchel inequality gives 
\begin{equation*}
u(k-B)\leq Z(k-B)+\Phi (Z)
\end{equation*}%
for every positive random variable $Z$ and $k\in L^{0}.$ Let $Q\in \mathcal{N%
}_{B}^{W}$ and take any $\lambda >0.$ By setting $Z=\lambda \frac{dQ^{r}}{dP}
$, fixing $k\in K_{B}^{W}$ and taking expectations we have 
\begin{equation*}
E[u(k-B)]\leq \lambda E_{Q^{r}}[k-B]+E\left[ \Phi \left( \lambda \frac{dQ^{r}%
}{dP}\right) \right] .
\end{equation*}%
>From the definition of $K_{B}^{W}$ 
\begin{equation*}
E_{Q^{r}}[k]\leq \widehat{Q^{s}}(-B)+\Vert Q^{s}\Vert
\end{equation*}%
whence 
\begin{equation*}
E[u(k-B)]\leq \lambda (\widehat{Q}(-B)+\Vert Q^{s}\Vert )+E\left[ \Phi
\left( \lambda \frac{dQ^{r}}{dP}\right) \right] .
\end{equation*}%
The expression $E\left[ \Phi \left( \lambda \frac{dQ^{r}}{dP}\right) \right] 
$ may be equal to $+\infty ,$ but for each $Q\in \mathcal{N}_{B}^{W}$ there
is a positive $\lambda $ for which it is finite. The thesis then follows.
\end{proof}

\subsection{The main duality result}

\begin{theorem}
\label{conj1} Fix a loss variable $W$ and a liability $B\in \mathcal{A}_{u}$%
. If 
\begin{equation}
\sup_{H\in \mathcal{H}^{W}}E\left[ u\left( \int_{0}^{T}H_{t}dS_{t}-B\right) %
\right] <u(+\infty )  \label{finite_condition}
\end{equation}%
then $\mathcal{N}_{B}^{W}$ is not empty and 
\begin{eqnarray}
\sup_{H\in \mathcal{H}^{W}} &&\hspace{-0.3in}E\left[ u\left(
\int_{0}^{T}H_{t}dS_{t}-B\right) \right] =U_{B}^{W}  \notag \\
&=&\min_{\lambda >0,\text{ }Q\in \mathcal{N}_{B}^{W}}\left\{ \lambda 
\widehat{Q}(-B)+E\left[ \Phi \left( \lambda \frac{{dQ}^{r}}{{dP}}\right) %
\right] +\lambda \Vert Q^{s}\Vert \right\} .  \label{genut}
\end{eqnarray}%
The minimizer $\lambda _{B}$ is unique, while the minimizer $Q_{B}$ is
unique only in the regular part $Q_{B}^{r}$.

Suppose in addition that the utility $u$ satisfies assumption $A.$ Then, 
\begin{equation}
U_{B}^{W}=E[u(f_{B}-B)],  \label{optimal_claim_imp}
\end{equation}%
where the unique maximizer is 
\begin{equation}
f_{B}=\left( -\Phi ^{\prime }(\lambda _{B}\frac{dQ_{B}^{r}}{dP})+B\right)
\in K_{B}^{W}  \label{optimal_claim_exp}
\end{equation}%
and satisfies 
\begin{equation}
E_{Q_{B}^{r}}[f_{B}]=\widehat{Q_{B}^{s}}(-B)+\Vert Q_{B}^{s}\Vert .
\label{budget}
\end{equation}
\end{theorem}

\begin{proof}
First observe that it follows from (\ref{step1}) that 
\begin{equation*}
\sup_{H\in \mathcal{H}^{W}}E\left[ u\left( \int_{0}^{T}H_{t}dS_{t}-B\right) %
\right] =\sup_{k\in K^{W}}E[u(k-B)]=\sup_{f\in C^{W}}E[u(f-B)].
\end{equation*}%
Moreover, Lemma \ref{continuity} enables the application of Fenchel duality
theorem to get 
\begin{eqnarray*}
\sup_{f\in C^{W}}E[u(f-B)] &=&\sup_{f\in C^{W}}I_{u}^{B}(f)=\min_{Q\in
(C^{W})^{0}}J_{u}^{B}(Q) \\
&=&\min_{Q\in (C^{W})^{0}}\left\{ E[\Phi (Q^{r})]+\widehat{Q}(-B)+\Vert
Q^{s}\Vert \right\}
\end{eqnarray*}%
where the last equality is guaranteed by Lemma \ref{LemmaJJ}. Now if the
optimal $Q$ had $Q^{r}=0$, then we would have 
\begin{equation*}
\sup_{f\in C^{W}}E[u(f-B)]=\Phi (0)+\widehat{Q}^{s}(-B)+\Vert Q^{s}\Vert
\geq u(+\infty ),
\end{equation*}%
since $\widehat{Q^{s}}(-B)+\Vert Q^{s}\Vert \geq 0$, according to (\ref%
{singineq}), and $\Phi (0)=u(\infty )$. Because this contradicts condition (%
\ref{finite_condition}), $Q^{r}\neq 0$ and a re-parametrization of the
domain of minimization in terms of $\mathcal{N}_{B}^{W}$ leads to 
\begin{equation*}
\sup_{f\in C^{W}}E[u(f-B)]=\min_{\lambda >0,\text{ }Q\in \mathcal{N}%
_{B}^{W}}\left\{ \lambda \widehat{Q}(-B)+E\left[ \Phi \left( \lambda \frac{{%
dQ}^{r}}{{dP}}\right) \right] +\lambda \Vert Q^{s}\Vert \right\} .
\end{equation*}%
Uniqueness of $\lambda _{B}$ and $Q_{B}^{r}$ follow from strict convexity of
the dual objective function in $\lambda $ and $Q^{r}$. However, the
dependence of the dual objective function on $Q^{s}$ is mixed: it is linear
in the norm $\Vert \cdot \Vert $-part due to \eqref{Q} (see \cite[%
Proposition 10]{BiaFri06} and generally convex in the term $\widehat{Q^{s}}%
(-B)$, although this term may also reduce to a linear one in the special
case $B\in L^{\widehat{u}}$. Therefore, the optimal singular functional
might not be unique. Thanks to Lemma \ref{ineq}, the equalities 
\begin{equation*}
\sup_{k\in K^{W}}E[u(f-B)]=U_{B}^{W}=\min_{\lambda >0,\text{ }Q\in \mathcal{N%
}_{B}^{W}}\left\{ \lambda \widehat{Q}(-B)+E\left[ \Phi \left( \lambda \frac{{%
dQ}^{r}}{{dP}}\right) \right] +\lambda \Vert Q^{s}\Vert \right\}
\end{equation*}%
are immediate. Under assumption $A,$ the expression for $f_{B}$ can be
derived by observing that any minimizer $Q^{B}$ is obtained as the minimizer
of 
\begin{equation*}
\min_{Q\in \mathcal{N}_{B}^{W}}\left\{ \lambda _{B}\widehat{Q}(-B)+E\left[
\Phi \left( \lambda _{B}\frac{{dQ}^{r}}{{dP}}\right) \right] +\lambda
_{B}\Vert Q^{s}\Vert \right\}
\end{equation*}%
and from a standard combination of the results in Propositions \ref{th111}
and \ref{perp}.
\end{proof}

\begin{corollary}
\label{B} Whenever $B\in \mathcal{B}=\mathcal{A}_{u}\cap L^{\widehat{u}}$ we
have that $\widehat{Q}(-B)=-Q(B)$ in \eqref{genut} and $\mathcal{N}_{B}^{W}=%
\mathcal{N}^{W}$. Moreover, if both $W$ and $B$ are in $M^{\widehat{u}}$,
then $\mathcal{N}_{B}^{W}$ can be replaced by the set $\mathbb{M}_{\sigma
}\cap \mathcal{L}_{\Phi }$ of $\sigma $- martingale probabilities with
finite generalized entropy and no singular term appears in \eqref{genut}.
\end{corollary}

\begin{proof}
The first statement is clear from the definition of $\widehat{Q}$. For the
second statement, notice that when $W\in M^{\widehat{u}}$ then the regular
component $Q^{r}$ of $Q\in \mathcal{M}^{W}$ is already in $\mathcal{M}^{W}$
(see \cite[Lemma 41]{BiaFri06}). If $B$ is in $M^{\widehat{u}}$ as well,
then $Q_{s}(B)=0$. Since $\Vert Q^{s}\Vert \geq 0$, the minimum must be
achieved on the set $\left\{ Q^{r}\mid Q\in \mathcal{N}_{B}^{W}\right\} =%
\mathbb{M}_{\sigma }\cap \mathcal{L}_{\Phi }$.
\end{proof}

\bigskip

We can see from \eqref{genut} that the singular part $Q^{s}$ in the dual
objective function plays a double role. Its norm $\Vert Q^{s}\Vert $ sums up
the generic risk of the high exposure in the market generated by $S$. When
the agent sells $B$, there is obviously an extra idiosyncratic exposure.
Given our very general assumptions on $B$, this extra exposure may also be
extremely risky, and this is expressed by the term $\widehat{Q^{s}}(-B)$. Of
course, the presence of \textquotedblleft high exposure" terms in the dual
does not imply that the actual minimizer $Q_{B}$ \emph{must} have a non-zero
singular part. However, in the next section we construct some examples
displaying the more interesting situation where $Q_{B}^{s}$ is necessarily
non-zero. In view of \eqref{budget}, a sufficient condition for this is $%
E_{Q_{B}^{r}}[f_{B}]>0$. The condition is by no means necessary, since it
could happen that $Q_{B}^{s}\neq 0$ but $\Vert Q_{B}^{s}\Vert +\widehat{%
Q_{B}^{s}}(-B)=0$ in \eqref{budget}.

It is interesting to investigate and possibly derive more accurate bounds
for $\widehat{Q^s}(-B)$. The next Proposition gives\emph{\ a priori} good
bounds for this singular contribution.

\begin{proposition}
For any $B\in \mathcal{A}_{u}$ set 
\begin{equation*}
L=\sup \{\beta >0\mid E[\widehat{u}(\beta B^{+})]<+\infty \}
\end{equation*}%
and fix any $Q\in \mathcal{N}_{B}^{W}$. Then 
\begin{equation}
\widehat{Q^{s}}(-B)\geq -\frac{1}{L}\Vert Q^{s}\Vert .  \label{b1}
\end{equation}%
If $B^{-}$ is also in $L^{\widehat{u}}$, set 
\begin{equation*}
l=\sup \{\alpha >0\mid E[\widehat{u}(\alpha B^{-})]<+\infty \}.
\end{equation*}%
Then 
\begin{equation}
-\frac{1}{L}\Vert Q^{s}\Vert \leq {Q^{s}}(-B)\leq \frac{1}{l}\Vert Q^{s}\Vert
\label{bounds}
\end{equation}%
and in particular we recover again $Q^{s}(B)=0$ when $B\in M^{\widehat{u}}$.
\end{proposition}

\begin{proof}
>From (\ref{e}), $E[u(-(1+\varepsilon)B^+ )] <+\infty $, so $L\geq 1+
\epsilon $. For any $b<L$, \eqref{singineq} gives $\|Q^s\|\geq b \, Q^s(B^+) 
$ and therefore 
\begin{equation*}
\widehat{Q^s}(-B)\geq - Q^s( B^+) \geq -\frac{1}{b } \|Q^s\|
\end{equation*}
whence the desired $\widehat{Q^s}(-B) \geq -\frac{1}{L } \|Q^s\| $. To prove
the right inequality in \eqref{bounds}, observe that the additional
hypothesis on $B^-$ means $l>0$ and $-\alpha B^- \in \mathrm{Dom}(I_u)$ for
any $\alpha<l$. Hence 
\begin{equation*}
Q^s(-B)\leq Q^s(B^-) = \frac{1}{\alpha} Q^s(\alpha B^-)\leq \frac{1}{\alpha}%
\| Q^s\| \ \ \text{ for all } \alpha<l
\end{equation*}
\end{proof}

The result in Theorem \ref{conj1} does not guarantee in full generality that
the optimal random variable $f_{B}\in K_{B}^{W}$ can be represented as
terminal value from an investment strategy in $L(S)$, that is, $%
f_{B}=\int_{0}^{T}H_{t}dS_{t}$. The next proposition presents a partial
result in this direction.

\begin{proposition}
\label{hedge}Suppose that the utility $u$ satisfies assumption $A.$ Under
the same hypotheses of Theorem \ref{conj1}, if $Q_{B}^{s}=0$ and $%
Q_{B}^{r}\sim P$, then $f_{B}$ can be represented as terminal wealth from a
suitable strategy $H$. \label{representation}
\end{proposition}

\begin{proof}
It follows from Theorem \ref{conj1} that $f_{B}$ must satisfy 
\begin{equation*}
E_{Q^{r}}[f_{B}]\leq \widehat{Q^{s}}(-B)+\Vert Q^{s}\Vert \ \ \forall Q\in 
\mathcal{N}_{B}^{W}
\end{equation*}%
and equality must hold at any optimal $Q_{B}$, according to \eqref{budget}.
When the optimal $Q_{B}$ has zero singular part, then it is a $\sigma $%
-martingale measure with finite entropy, according to \eqref{measures}. This
being the case, it is easy to see that the dual problem could be
reformulated as a minimum over $M_{\sigma }\cap \mathcal{L}_{\Phi }$. In
this simplified setup, one can show exactly as in \cite[Therem 4, Theorem 1
(d)]{biafri05} that the optimal $f_{B}$ belongs in fact to 
\begin{equation*}
\bigcap\limits_{Q\in M_{\sigma }\cap \mathcal{L}_{\Phi }}\{f\mid f\in L^{1}{Q%
},E_{Q}[f]\leq 0\}
\end{equation*}%
and can be represented as terminal wealth from a suitable strategy $H$.
\end{proof}

\vspace{0.2in}

\subsection{Exponential utility}

For an exponential utility function $u(x)=-e^{-\gamma x},$ $\gamma >0$, we
have 
\begin{eqnarray*}
\Phi (y) &=&\frac{y}{\gamma }\log \frac{y}{\gamma }-\frac{y}{\gamma } \\
\widehat{u}(x) &=&e^{\gamma |x|}-1
\end{eqnarray*}%
Using (\ref{uu}), we see that in this case $M^{\widehat{u}}$ consists of
those random variables that have \emph{all} the (absolute) exponential
moments finite, while the larger space $L^{\widehat{u}}$ corresponds to
random variables that have \emph{some} finite exponential moment.

Moreover, since $\widehat{\Phi }(y)=(\Phi (|y|)-\Phi (\gamma
))I_{\{|y|>\gamma \}}$ and $\Phi (0)<\infty $, we have that 
\begin{equation}
E[\widehat{\Phi }(f)]<\infty \quad \Longleftrightarrow \quad E[\Phi
(|f|)]<\infty .
\end{equation}%
Finally, since $\widehat{\Phi }$ in this case satisfies the $\Delta _{2}$%
--growth condition (see \cite[pp 22, 77]{RR}), the subspace $M^{\widehat{%
\Phi }}$ coincides with $L^{\widehat{\Phi }}$, that is, $E[\widehat{\Phi }%
(\alpha f)]<\infty $ for \emph{some} $\alpha >0$ if and only if $E[\widehat{%
\Phi }(\alpha f)]<\infty $ for \emph{all} $\alpha >0$.

The duality result for an exponential utility, which clearly satisfies
Assumption (A), follows directly as a corollary of our main Theorem \ref%
{conj1}. Since $u(\infty )=0,$ the condition (\ref{gains}) automatically
holds for all $\mathcal{F}_{T}$ measurable random variables $B$ \ and
furthermore, 
\begin{equation*}
\mathcal{L}_{\Phi }=\{Q\text{ probab},\text{ }Q\ll P\mid E\left[ \Phi \left( 
\frac{dQ}{dP}\right) \right] <+\infty \}=\{Q\text{ probab},\text{ }Q\ll
P\mid E\left[ \widehat{\Phi }\left( \frac{dQ}{dP}\right) \right] <+\infty \}.
\end{equation*}

\begin{corollary}
\label{cor1} Suppose that the random endowment $B\in L^{0}(\Omega ,\mathcal{F%
}_{T},P)$ satisfies%
\begin{equation*}
E[e^{\gamma (1+\epsilon )B}]<+\infty \text{ for some }\epsilon >0
\end{equation*}%
and suppose that there exists a loss variable $W$ satisfying 
\begin{equation}
\sup_{H\in \mathcal{H}^{W}}E\left[ -e^{-\gamma (\int_{0}^{T}H_{t}dS_{t}-B)}%
\right] <0.  \label{H0}
\end{equation}%
Then $\mathcal{N}_{B}^{W}$ is not empty and 
\begin{eqnarray}
\sup_{H\in \mathcal{H}^{W}} &&\hspace{-0.3in}E\left[ -e^{-\gamma
(\int_{0}^{T}H_{t}dS_{t}-B)}\right] =  \notag \\
&&-\exp \left\{ -\min_{Q\in \mathcal{N}_{B}^{W}}\left( H(Q^{r}|P)+\gamma 
\widehat{Q}(-B)+\gamma \Vert Q^{s}\Vert \right) \right\} ,  \label{hh}
\end{eqnarray}%
where $H(Q^{r}|P)=E\left[ \frac{dQ^{r}}{dP}\log \left( \frac{dQ^{r}}{dP}%
\right) \right] $ denotes the relative entropy of $Q^{r}$ with respect to $P$%
. The minimizer $Q_{B}\in \mathcal{N}_{B}^{W}$ is unique only in the regular
part $Q_{B}^{r}$. In addition, 
\begin{equation*}
\sup_{H\in \mathcal{H}^{W}}E\left[ -e^{-\gamma (\int_{0}^{T}H_{t}dS_{t}-B)}%
\right] =E[-e^{-\gamma (f_{B}-B)}],
\end{equation*}%
where the optimal claim is 
\begin{equation*}
f_{B}=-\frac{1}{\gamma }\ln \left( \frac{\lambda _{B}}{\gamma }\frac{%
dQ_{B}^{r}}{dP}\right) +B,
\end{equation*}%
where $\lambda _{B}=\gamma \exp (H(Q_{B}^{r}|P)+\gamma \widehat{Q_{B}}%
(-B)+\gamma \Vert Q_{B}^{s}\Vert )=-\frac{1}{\gamma }U_{B}^{W}$, and it
satisfies

\begin{enumerate}
\item $f_{B}\in L^{1}(Q^{r})$, $E_{Q^{r}}[f_{B}]\leq \widehat{Q^{s}}%
(-B)+\Vert Q^{s}\Vert $ for all $Q\in \mathcal{N}_{B}^{W}$ (i.e. it belongs
to $K_{B}^{W}$)

\item $E_{Q_B^{r}}[f_B]= \widehat{Q_B^s} (-B)+ \Vert Q_B^{s}\Vert$
\end{enumerate}

Whenever $B$ has some exponential (absolute) moments finite, $\widehat{Q}%
(-B)=-Q(B)$. Also, if both $W$ and $B$ have all the exponential moments
finite, then $\mathcal{N}_{B}^{W}$ can be replaced by the \textquotedblleft
classic\textquotedblright\ set of probabilities $Q\in \mathbb{M}_{\sigma }$
that have finite relative entropy, i.e. $E[\frac{dQ}{dP}\ln (\frac{dQ}{dP}%
)]<+\infty $, and no singular term appears in (\ref{hh}).
\end{corollary}

\begin{proof}
The conditions on $B$ are exactly those in Theorem \ref{conj1}, adapted to
the exponential case. So, directly from Theorem \ref{conj1} 
\begin{eqnarray*}
\sup_{H\in \mathcal{H}^{W}} &&\hspace{-0.3in}E\left[-e^{-\gamma ( \int_0^T
H_t dS_t -B)}\right] = \\
&&\min_{\lambda >0,\text{ }Q\in \mathcal{N}_B^{W}}\left\{ \lambda \widehat
Q(-B)+ E\left[ \frac{\lambda }{\gamma }\frac{dQ^{r}}{dP}\log \left( \frac{%
\lambda }{\gamma }\frac{dQ^{r}}{dP}\right) -\frac{\lambda }{\gamma }\frac{%
dQ^{r}}{dP}\right] +\lambda \| Q^{s}\| )\right\} ,
\end{eqnarray*}%
and an explicit minimization over $\lambda >0$ leads to the duality formula %
\eqref{hh}. The remaining assertions follow as in the proof of Theorem \ref%
{conj1}.
\end{proof}

\subsubsection{Examples with nonzero singular parts}

We now explore the case of an exponential utility to construct two examples
where the existence of a nonzero singular part in the dual optimizer can be
asserted explicitly.

\begin{example}
Consider a one period model with $S_{0}=0$ and $S_{1}=YZ$ where $Y$ is an
exponential random variable with density $f(y)=e^{-y}$, $y\geq 0$ and $Z$ is
a discrete random variable taking the values $\{1,-\frac{1}{2},\ldots ,\frac{%
1}{n}-1,\ldots \}$. Assume that $Y$ and $Z$ are independent and let 
\begin{eqnarray*}
p_{1}:= &&P(Z=1)>0 \\
p_{n}:= &&P\left( Z=\frac{1}{n}-1\right) >0,\quad n\geq 2
\end{eqnarray*}%
be the probability distribution of $Z$. For an investor with exponential
utility $u(x)=-e^{-x}$, it is clear that the random variable $W=1+Y$ is
suitable and compatible. Suppose now that $B=\alpha (Y,Z)$, where $\alpha $
is a bounded Borel function, so that the seller of the claim $B$ faces the
problem 
\begin{equation*}
\sup_{h\in \mathbb{R}}E\left[ -e^{-hS_{1}+B}\right] =\sup_{h\in \mathbb{R}}E%
\left[ -e^{-hZY+\alpha (Y,Z)}\right] .
\end{equation*}

Because $Y$ is exponentially distributed with parameter $1$, $\alpha $ is
bounded, $-1<Z\leq 1$ and independent from $Y$, a necessary condition for
the expectation above to be finite is that $-1<h\leq 1$. Now the function 
\begin{equation*}
g(h)=E\left[ -e^{-hS_{1}+B}\right] ,
\end{equation*}%
has a formal derivative given by 
\begin{equation*}
g^{\prime }(h)=E\left[ S_{1}e^{-hS_{1}+B}\right] =p_{1}E\left[
Ye^{-hY+\alpha (Y,Z)}\right] +\sum_{n\geq 2}p_{n}z_{n}E\left[
Ye^{-hz_{n}Y+\alpha (Y,z_{n})}\right] .
\end{equation*}%
Since $-1<z_{n}<0$ for $n\geq 2$, we have that 
\begin{equation*}
g^{\prime }(h)\geq p_{1}E\left[ Ye^{-Y+B}\right] -\sum_{n\geq 2}p_{n}E\left[
Ye^{-z_{n}Y+\alpha (Y,z_{n})}\right] .
\end{equation*}%
When $p_{n}\rightarrow 0$ sufficiently fast, this expression is not only
well defined but strictly positive. Therefore, by adjusting the distribution
of $Z$, we can guarantee that $0<g^{\prime }(h)<\infty $ for all $-1<h\leq 1$%
. Therefore, the function $g(h)$ is strictly increasing and attains its
maximum at $h=1$. But this implies that 
\begin{equation*}
\sup_{h\in \mathbb{R}}E\left[ -e^{-hS_{1}+B}\right] =E\left[ -e^{-S_{1}+B}%
\right] ,
\end{equation*}%
so that the optimizer for the primal problem is $f_{B}=S_{1}$. From the
identity 
\begin{equation*}
u^{\prime }(f_{B}-B)=\lambda _{B}\frac{dQ_{B}^{r}}{dP},
\end{equation*}%
we obtain that the optimizer for the dual problem has a regular part given
by 
\begin{equation}
\frac{dQ_{B}^{r}}{dP}=\frac{e^{-S_{1}+B}}{E[e^{-S_{1}+B}]}.
\label{reg_example}
\end{equation}%
Using \eqref{reg_example} to calculate the expectation of $f_{B}$ with
respect to $Q_{B}^{r}$, we conclude from \eqref{budget} that 
\begin{equation*}
Q_{B}^{s}(-B)+\Vert Q_{B}^{s}\Vert =E_{Q_{B}^{r}}[f_{B}]=\frac{%
E[S_{1}e^{-S_{1}+B}]}{E[e^{-S_{1}+B}]}=\frac{g^{\prime }(1)}{E[e^{-S_{1}+B}]}%
>0,
\end{equation*}%
which implies that $Q_{B}^{s}\neq 0$.

Observe that a proper selection of the probabilities $p_{n}$ also guarantees
that setting $B=0$ in the expressions above does not alter the domain of the
function $g(h)$ and the remaining calculations. In particular, the maximum
of $E[-e^{-hS_{1}}]$ would be still attained at $h=1$, which implies that
the optimizers $f_{0}$ and $f_{B}$ for the primal problem with and without
the claim coincide. This means that the investor does not use the underlying
market to hedge the claim, despite the fact that $B=\alpha (Y,Z)$ is
explicitly correlated with $S_{1}=YZ$. Such behavior stems from the fact
that the risk associated with the unboundedness of the underlying outweighs
the risk associated with the bounded claim. This should be contrasted with
the case of locally bounded markets, where even a bounded claim leads to a
different optimizer for the primal problem.
\end{example}

\begin{example}
Consider now the same setting as in the previous example, but with a claim
of the form $B=\delta Y$, $0<\delta <1$, so that the investor faces the
problem 
\begin{equation*}
\sup_{h\in \mathbb{R}}E\left[ -e^{-(hS_{1}-\delta Y)}\right] =\sup_{h\in 
\mathbb{R}}E\left[ -e^{-(hZ-\delta )Y}\right] .
\end{equation*}%
A necessary condition for the expectation above to be finite is $-(1-\delta
)<h\leq (1-\delta )$, since $-1<Z\leq 1$. Define the function 
\begin{equation*}
g(h)=E\left[ -e^{-hS_{1}+\delta Y}\right] ,
\end{equation*}%
with derivative 
\begin{equation*}
g^{\prime }(h)=E\left[ S_{1}e^{-hS_{1}+\delta Y}\right] =p_{1}E\left[
Ye^{-(h-\delta )Y}\right] +\sum_{n\geq 2}p_{n}z_{n}E\left[
Ye^{-(hz_{n}-\delta )Y}\right] .
\end{equation*}%
As before, 
\begin{equation*}
g^{\prime }(h)\geq p_{1}E\left[ Ye^{-(1-\delta )Y}\right] -\sum_{n\geq
2}p_{n}E\left[ Ye^{-(z_{n}+\delta )Y}\right] ,
\end{equation*}%
which can be made strictly positive for $p_{n}\rightarrow 0$ sufficiently
fast (as a consequence, we can assume $p_{1}\gg p_{n}$). Therefore, $%
0<g^{\prime }(h)<\infty $ for all $-(1-\delta )<h\leq (1-\delta )$ and the
function $g(h)$ attains its maximum at $h=1-\delta $. We then obtain that $%
f_{B}=(1-\delta )S_{1}$, which implies that 
\begin{equation}
\frac{dQ_{B}^{r}}{dP}=\frac{e^{-(1-\delta )S_{1}+\delta Y}}{E[e^{-(1-\delta
)S_{1}+\delta Y}]},  \label{reg_example_2}
\end{equation}%
in view of the identity 
\begin{equation*}
u^{\prime }(f_{B}-B)=\lambda _{B}\frac{dQ_{B}^{r}}{dP}.
\end{equation*}%
As before, inserting this in \eqref{budget} 
\begin{equation*}
Q_{B}^{s}(-B)+\Vert Q_{B}^{s}\Vert =E_{Q_{B}^{r}}[f_{B}]=\frac{E[(1-\delta
)S_{1}e^{-(1-\delta )S_{1}+\delta Y}]}{E[e^{-(1-\delta )S_{1}+\delta Y}]}=%
\frac{g^{\prime }(1)}{E[e^{-(1-\delta )S_{1}+\delta Y}]}>0,
\end{equation*}%
which implies that $Q_{B}^{s}\neq 0$.

Apart from the appearance of a nonzero singular part in the pricing measure,
an interesting feature of this example is the excess hedge $%
f_{B}-f_{0}=-\delta S_{1}$ induced by the presence of the claim $B$. Observe
that the selection $p_{1}\gg p_{n}$ guarantees that $B$ is positively
correlated with $S_{1}$, since 
\begin{equation*}
\mbox{Cov}(B,S_{1})=\delta E[Z]\mbox{Var}\lbrack Y],
\end{equation*}%
and $E[Z]$ is positive when $p_{1}$ is sufficiently larger than $p_{n}$.
This would suggest that the seller of $B$ should hedge it by \emph{buying}
more shares of $S$. What our analyses indicates is that this intuition is in
fact \emph{wrong}, since the excess hedge due to the presence of $B$
consists of \emph{selling} $\delta $ shares of $S$. The explanation for this
counterintuitive result relies on the fact that $B$ is not \emph{perfectly}
correlated with $S$. In fact, whenever $Z<0$, the risks of large downward
moves in $S_{1}=YZ$ and large upward moves in $B=\delta Y$ are both related
to the same exponential random variable $Y$. Therefore, in the presence of $%
B $, the preference structure prohibits to buy more than $1-\delta $ shares,
which must be then the new optimum.
\end{example}

\section{The indifference price $\protect\pi \label{sec4}$}

\subsection{Definition and domain of $\protect\pi $}

Consider an agent with utility $u$ (not necessarily satisfying Assumption
(A)), initial endowment $x$ and investment possibilities given by $\mathcal{H%
}^{W}$ who seeks to sell a claim $B$. As pointed out in Section \ref%
{introduction}, the indifference price $\pi (B)$ for this claim is defined
as the implicit solution to \eqref{def_indiff}. In view of the duality
result of Theorem \ref{conj1}, we now rephrase this definition in terms of
the function 
\begin{equation}
U_{B}^{W}(x):=\sup_{k\in K^{W}}E[u(x+k-B)].  \label{primal_B_2}
\end{equation}%
Comparing this with \eqref{genut}, we see that the optimal value $%
U_{B}^{W}(0)$ is exactly what has been there denoted by $U_{B}^{W}$. Notice
that we could alternatively denote \eqref{primal_B_2} by $U_{B-x}^{W}$,
which would be consistent with \eqref{genut} for a claim of the form $(B-x)$%
. We prefer $U_{B}^{W}(x)$ instead, since it better illustrates the
different financial roles played by the initial endowment $x$ and the claim $%
B$.

\begin{definition}
Provided that the related maximization problems are well--posed, the
seller's indifference price $\pi(B)$ of the claim $B$ is the implicit
solution of the equation 
\begin{eqnarray}  \label{indiffprice}
U_0^W(x) =U^W_{B}(x+\pi(B))
\end{eqnarray}
that is, $\pi(B)$ is the additional initial money that makes the optimal
utility with the liability $B$ equal to the optimal utility without $B$.
\end{definition}

The next lemma shows that the class 
\begin{equation}
\mathcal{B}=\mathcal{A}_{u}\cap L^{\widehat{u}}=\{B\in L^{\widehat{u}}\mid E[%
\widehat{u}((1+\epsilon )B^{+})]<+\infty \text{ for some }\epsilon >0\}
\label{domain_B}
\end{equation}%
of claims $B,$ for which we compute indifference prices, is considerably
large and has desirable properties. Note that the equivalence \eqref{uu}
says that $E[\widehat{u}((1+\epsilon )B^{+})]<+\infty $ if and only if $B$
satisfies \eqref{e}, so that (\ref{bb}) and (\ref{domain_B}) agree. In other
words, $\mathcal{B}$ consists of the set of claims which, in addition to
satisfying the hypotheses of Theorem \ref{conj1}, are also in $L^{\widehat{u}%
}$. Upon fixing the loss variable $W$, the strengthening assumption $B\in L^{%
\widehat{u}}$ allows us to use Corollary \ref{B} and guarantees that the set
of dual functionals $\mathcal{N}_{B}^{W}$ does not depend on $B$ and reduces
to the set $\mathcal{N}^{W}$ defined in \eqref{ennew}.

\begin{lemma}
\label{domainrho} 
\begin{equation}  \label{BB}
\mathcal{B} = \{B\in L^{\widehat{u}}\mid (-B) \in\mathrm{\ int}(\mathrm{Dom}%
(I_{u}))]
\end{equation}
and therefore has the properties:

\begin{enumerate}
\item $\mathcal{B}$ is convex and open in $L^{\widehat{u}}$;

\item If $B_1 \in \mathcal{B}$ and $B_2\leq B_1$, then $B_2\in\mathcal{B} $.

\item $\mathcal{B}$ contains $M^{\widehat{u}}$ (and thus $L^{\infty}$);

\item for any given $B\in \mathcal{B}$ and $C\in M^{\widehat{u}} $, we have
that $B+C \in \mathcal{B}$. In particular, $B+c \in \mathcal{B}$ for all
constants $c\in\mathbb{R}$.
\end{enumerate}
\end{lemma}

\begin{proof}
As remarked after (\ref{e}), we already know that $B$ satisfies \eqref{e}
iff $-B^{+}\in \mathrm{int}(\mathrm{Dom}(I_{u}))$. Under the extra condition 
$B\in L^{\widehat{u}}$, $B$ satisfies \eqref{e} iff $-B\in \mathrm{int}(%
\mathrm{Dom}(I_{u}))$, which shows \eqref{BB}.

Then, $\mathcal{B}$ is obviously open and convex (property 1) and property 2
is a consequence of the monotonicity of $I_{u}$. It is evident that $M^{%
\widehat{u}}$ is contained in $\mathcal{B}$, since $C\in M^{\widehat{u}}$
iff $E[\widehat{u}(kC)]<+\infty $ for all $k>0$ (property 3). In order to
prove property 4, fix $B\in \mathcal{B}$ and a convenient $\epsilon $. For
any $C$ in $M^{\widehat{u}}$, set $r=\frac{\frac{\epsilon }{2}}{(1+\epsilon
)(1+\frac{\epsilon }{2})}$. Then 
\begin{equation*}
E\left[ \widehat{u}\left( (1+\frac{\epsilon }{2})(B+C)^{+}\right) \right]
\leq \frac{1+\frac{\epsilon }{2}}{1+\epsilon }E\left[ \widehat{u}%
((1+\epsilon )B^{+})\right] +\frac{\epsilon /2}{1+\epsilon }E\left[ \widehat{%
u}\left( \frac{C^{+}}{r}\right) \right] <+\infty .
\end{equation*}
\end{proof}

\subsection{The properties of $\protect\pi$}

The next Proposition lists the various properties of the \emph{indifference
price functional} $\pi $, defined on the set $\mathcal{B}\subseteq L^{%
\widehat{u}}$. Some results are new, in particular the regularity of the map
and the description of the conjugate $\pi ^{\ast }$ and of the
subdifferential $\partial \pi $. They are nice consequences of the choice of
the natural Orlicz framework and the proofs are quite short and easy. The
other items are extensions of well--established results to the present
general setup (see e.g. \cite{bec} or the recent \cite[Prop. 7.5]{oz} and
the references therein). A recent reference book for the necessary notions
from Convex Analysis is \cite{bz}.

In the next proposition, the assumption that $U_{0}^{W}(x)<u(+\infty )$ can
be replaced by $\mathcal{N}^{W}\neq \emptyset ,$ whenever the utility
function satisfies assumption (A). Indeed, in this case Proposition \ref%
{perp} and $\mathcal{N}^{W}\neq \emptyset $ guarantees that $%
U_{0}^{W}(x)<u(+\infty )$ for all $x\in \mathbb{R}$.

\begin{proposition}
\label{propindiff}Fix a loss variable $W$ and an initial wealth $x\in 
\mathbb{R}$ such that $U_{0}^{W}(x)<u(+\infty )$. The seller's indifference
price 
\begin{equation*}
\pi :\mathcal{B}\rightarrow \mathbb{R}
\end{equation*}%
verifies the following properties:

\begin{enumerate}
\item $\pi $ \textbf{is well--defined}. The solution to the equation %
\eqref{indiffprice} above exists and it is unique.

\item \textbf{Convexity and monotonicity.} $\pi$ is a convex, monotone
non--decreasing functional.

\item \textbf{Translation invariance.} Given $B\in \mathcal{B}$, $\pi(B+c)=
\pi(B)+c$ for any $c \in \mathbb{R}$.

\item \textbf{Regularity.} $\pi$ is norm continuous and subdifferentiable.

\item \textbf{Dual representation. } ${\pi }$ admits the representation 
\begin{equation}
\pi (B)=\max_{Q\in \mathcal{N}^{W}}\left( Q(B)-\alpha (Q)\right)   \label{pi}
\end{equation}%
where the (minimal) penalty term $\alpha (Q)$ is given by 
\begin{equation*}
\alpha (Q)=x+\Vert Q^{s}\Vert +\inf_{\lambda >0}\left\{ \frac{E[\Phi
(\lambda \frac{dQ^{r}}{dP})]-U_{0}^{W}(x)}{\lambda }\right\} .
\end{equation*}%
As a consequence, the subdifferential $\partial \pi (B)$ of $\pi $ at $B$ is
given by 
\begin{equation}
\partial \pi (B)=\mathcal{Q}_{B}^{W}(x+\pi (B))  \label{subdiff}
\end{equation}%
where $\mathcal{Q}_{B}^{W}(x+\pi (B))$ is the set of minimizers of the dual
problem associated with the right--hand side of \eqref{indiffprice}.

\item \textbf{Bounds.} $\pi $ satisfies the bounds 
\begin{equation*}
\max_{Q\in \mathcal{Q}_{0}^{W}(x)}Q(B)\leq \pi (B)\leq \sup_{Q\in \mathcal{N}%
^{W}}Q(B)
\end{equation*}%
If $W\in M^{\widehat{u}}$ and $B\in M^{\widehat{u}}$, the bounds above
simplify to 
\begin{equation*}
E_{Q^{\ast }}[B]\leq \pi (B)\leq \sup_{Q\in M_{\sigma }\cap \mathcal{L}%
_{\Phi }}E_{Q}[B]
\end{equation*}%
where the probability ${Q}^{\ast }\in M_{\sigma }\cap \mathcal{L}_{\Phi }$
is the unique dual minimizer in $\mathcal{Q}_{0}^{W}(x)$.

\item \textbf{\ Volume asymptotics.} For any $B\in \mathcal{B}$ we have 
\begin{equation}
\lim_{b\downarrow 0}\frac{\pi (bB)}{b}=\max_{Q\in \mathcal{Q}%
_{0}^{W}(x)}Q(B).  \label{betazero}
\end{equation}%
If $B$ is in $M^{\widehat{u}}$,

\begin{equation}
\lim_{b\rightarrow +\infty }\frac{\pi (bB)}{b}=\sup_{Q\in \mathcal{N}%
^{W}}Q(B).  \label{betastar}
\end{equation}%
If $W\in M^{\widehat{u}}$ and $B\in M^{\widehat{u}}$, the two volume
asymptotics above become 
\begin{equation*}
\lim_{b\downarrow 0}\frac{\pi (bB)}{b}=E_{Q^{\ast }}[B],\ \ \ \
\lim_{b\rightarrow +\infty }\frac{\pi (bB)}{b}=\sup_{Q\in M_{\sigma }\cap 
\mathcal{L}_{\Phi }}E_{Q}[B]
\end{equation*}%
where the probability ${Q}^{\ast }\in M_{\sigma }\cap \mathcal{L}_{\Phi }$
is the unique dual minimizer in $\mathcal{Q}_{0}^{W}(x)$.

\item \textbf{Price of replicable claims. } If $B \in \mathcal{B}$ is
replicable in the sense that $B = c + \int_0^T H_t dS_t $ with $H\in 
\mathcal{H}^W $, but also $-H \in \mathcal{H}^W$, then $\pi(B)=c$.
\end{enumerate}
\end{proposition}

\begin{proof}
Applying Theorem \ref{supmin}, eq. (\ref{ee}), we preliminary observe that
the assumption $U_{0}^{W}(x)<u(+\infty )$ implies $\mathcal{N}^{W}\neq
\varnothing $.

\begin{enumerate}
\item Let $F(p):=U_{B}^{W}(x+p).$ By standard arguments it can be shown that 
$F:\mathbb{R}\rightarrow (-\infty ,u(+\infty )]$ is concave and monotone
non--decreasing, though not necessarily strictly increasing. By monotone
convergence we also have 
\begin{equation}
\lim_{p\rightarrow +\infty }F(p)=u(+\infty ).  \label{lim12}
\end{equation}%
We now show that $\lim_{p\rightarrow -\infty }F(p)=-\infty ,$ so that $F(p)$
is not constantly equal to $u(+\infty )$. Fix $Q\in \mathcal{N}^{W}$ and
take $\lambda >0$ for which $E[\Phi (\lambda \frac{dQ^{r}}{dP})]$ is finite.
>From the inclusion $K^{W}\subseteq K_{B}^{W},$ proved in Lemma \ref{ineq},
and Fenchel inequality it follows, as in the second part of Lemma \ref{ineq}%
, that for all $k\in K^{W}$%
\begin{eqnarray*}
E[u(x+p+k-B)] &\leq &E\left[ (x+p+k-B)\lambda \frac{dQ^{r}}{dP}\right] +E%
\left[ \Phi \left( \lambda \frac{dQ^{r}}{dP}\right) \right] \\
&\leq &\lambda (x+p-Q(B)+\Vert Q^{s}\Vert )+E\left[ \Phi \left( \lambda 
\frac{dQ^{r}}{dP}\right) \right] <+\infty ,
\end{eqnarray*}%
so that the r.h.s. does not depend on $k$ anymore. Taking the $sup$ over $k$ 
\begin{equation*}
F(p)=U_{B}^{W}(x+p)\leq \lambda (x+p-Q(B)+\Vert Q^{s}\Vert )+E\left[ \Phi
\left( \lambda \frac{dQ^{r}}{dP}\right) \right]
\end{equation*}%
and then, passing to the limit for $p\rightarrow -\infty $, one obtains $%
\lim_{p\rightarrow -\infty }F(p)=-\infty $. The well-posedness of the
definition of $\pi $ is now straightforward. In fact, let $p_{L}$ be the
infimum of the set $\{p\in \mathbb{R}\mid F(p)=F(+\infty )=u(+\infty )\}$.
>From concavity, on $(-\infty ,p_{L})$ $F$ is continuous and strictly
monotone and thus a bijection onto the image $(-\infty ,u(+\infty ))$. Since 
$U_{0}^{W}(x)<u(+\infty )$, there always exists a unique $p$ such that $%
F(p)=U_{0}^{W}(x)$, namely the indifference price $\pi (B)$.

\item Convexity and monotonicity are consequences of the definition %
\eqref{indiffprice}, of the concavity and monotonicity of $u$ and of the
convexity of $\mathcal{H}^{W}$.

\item Translation invariance follows directly from the definition %
\eqref{indiffprice}.

\item For this item, observe that $\pi $ is a real valued, convex, monotone
functional on the convex open subset $\mathcal{B}$ of the Banach lattice $L^{%
\widehat{u}}$. It then follows from item 2 of Lemma \ref{domainrho} that the
extension $\widetilde{\pi }$ of $\pi $ on $L^{\widehat{u}}$ with the value $%
+\infty $ on $L^{\widehat{u}}\backslash \mathcal{B}$ is still monotone,
convex and translation invariant. Trivially, the interior of the proper
domain of $\widetilde{\pi }$ coincides with $\mathcal{B}$. Therefore, norm
continuity and sub-differentiability of $\widetilde{\pi }$ (and thus of $\pi 
$) on $\mathcal{B}$ follow from an extension of the classic Namioka-Klee
theorem for convex monotone functionals (see \cite{ruz}, but also \cite%
{BiaFri07} and \cite{cl} in the context of Risk Measures). As a consequence, 
$\pi $ admits a dual representation on $\mathcal{B}$ as 
\begin{equation}
\pi (B)=\widetilde{\pi }(B)=\max_{Q\in (L^{\widehat{u}})_{+}^{\ast },Q(%
\mathbf{1}_{\Omega })=1}\left\{ Q(B)-\pi ^{\ast }(Q)\right\}  \label{rho}
\end{equation}%
where $\pi ^{\ast }$ is the convex conjugate of $\widetilde{\pi }$, that is $%
\pi ^{\ast }:(L^{\widehat{u}})^{\ast }\rightarrow (-\infty ,+\infty ]$, 
\begin{equation*}
\pi ^{\ast }(z)=\sup_{B^{\prime }\in L^{\widehat{u}}}\{z(B^{\prime })-%
\widetilde{\pi }(B^{\prime })\}=\sup_{B\in \mathcal{B}}\{z(B)-{\pi }(B)\}.
\end{equation*}%
The normalization condition $Q(\mathbf{1}_{\Omega })=1$ in \eqref{rho}
derives from the translation invariance property. The subdifferential of $%
\pi $ at $B$ is, as always, given by 
\begin{equation}
\partial \pi (B)=\text{argmax}\{{Q(B)-\pi ^{\ast }(Q)}\}.  \label{subdiffe}
\end{equation}%
Note that, since $\pi (0)=0,$ $\pi ^{\ast }$ is nonnegative and thus it can
be interpreted as a penalty function. The next item presents a
characterization of $\pi ^{\ast }$ and therefore of $\partial \pi (B)$.

\item A dual representation for $\pi $ has just been obtained in \eqref{rho}%
. The current item is proved in two steps: first, we establish
representation \eqref{pi} with the penalty $\alpha $; second, we prove that $%
\alpha =\pi ^{\ast }$, that is $\alpha $ is the \emph{minimal penalty }%
function, which together with \eqref{subdiffe} gives \eqref{subdiff} and
completes the proof.\newline
Step 1. From the definition of $\pi (B)$ and from the dual formula %
\eqref{genut} 
\begin{eqnarray*}
U_{0}^{W}(x) &=&U_{B}^{W}(x+\pi (B)) \\
&=&\min_{\lambda >0,{Q}\in \mathcal{N}^{W}}\left\{ \lambda Q(-B+x+\pi
(B))+\lambda \Vert Q^{s}\Vert +E\left[ \Phi \left( \lambda \frac{dQ^{r}}{dP}%
\right) \right] \right\} .
\end{eqnarray*}%
Necessarily then 
\begin{equation*}
\pi (B)\geq Q(B)-\left[ x+\Vert Q^{s}\Vert +\frac{E[\Phi (\lambda \frac{%
dQ^{r}}{dP})]-U_{0}^{W}(x)}{\lambda }\right] \text{ for all }\lambda >0,Q\in 
\mathcal{N}^{W}
\end{equation*}%
and equality holds for the optimal $\lambda ^{\ast }$ and any ${Q}^{\ast
}\in \mathcal{Q}_{B}^{W}(x+\pi (B))$. Fixing $Q\in \mathcal{\ N}^{W}$ and
taking first the supremum over $\lambda >0$, we get 
\begin{equation*}
\pi (B)\geq Q(B)-\inf_{\lambda >0}\left[ x+\Vert Q^{s}\Vert +\frac{E[\Phi
(\lambda \frac{dQ^{r}}{dP})]-U_{0}^{W}(x)}{\lambda }\right] .
\end{equation*}%
Taking then the supremum over $Q$ we finally obtain 
\begin{equation*}
\pi (B)=\max_{Q\in \mathcal{N}^{W}}\left\{ Q(B)-\inf_{\lambda >0}\left[
x+\Vert Q^{s}\Vert +\frac{E[\Phi (\lambda \frac{dQ^{r}}{dP})]-U_{0}^{W}(x)}{%
\lambda }\right] \right\} 
\end{equation*}%
where equality holds for $\lambda ^{\ast },{Q}^{\ast }\in \mathcal{Q}%
_{B}^{W}(x+\pi (B))$. Observe that the following extension, still denoted by 
$\alpha $, 
\begin{equation*}
\alpha (Q)=\left\{ 
\begin{array}{c}
\inf_{\lambda >0}\left[ x+\Vert Q^{s}\Vert +\frac{E[\Phi (\lambda \frac{%
dQ^{r}}{dP})]-U_{0}^{W}(x)}{\lambda }\right] \text{ when }Q\in \mathcal{N}%
^{W} \\ 
+\infty \hfill \text{ otherwise} \\ 
\end{array}%
\right. 
\end{equation*}%
is $[0,+\infty ]$-valued and satisfies $\inf_{Q\in (L^{\widehat{u}})^{\ast
}}\alpha (Q)=0$. Therefore, it is a \emph{grounded} penalty function and
clearly 
\begin{equation*}
\pi (B)=\max_{Q\in (L^{\widehat{u}})_{+}^{\ast }}\left\{ Q(B)-\alpha
(Q)\right\} 
\end{equation*}%
and the set 
\begin{equation}
\mathrm{argmax}\left\{ Q(B)-\alpha (Q)\right\} \text{ coincides with }%
\mathcal{Q}_{B}^{W}(x+\pi (B)).  \label{subdiffer}
\end{equation}%
In particular, when $B=0$ 
\begin{equation}
\pi (0)=0\ \ \ \text{ and }\ \ \ \mathrm{argmax}\left\{ -\alpha (Q)\right\} =%
\mathrm{argmin}\left\{ \alpha (Q)\right\} =\mathcal{Q}_{0}^{W}(x).
\label{zero}
\end{equation}%
Step 2. As $\alpha $ provides another penalty function, a basic result in
convex duality ensures that $\pi ^{\ast }=\alpha ^{\ast \ast }$, i.e. $\pi
^{\ast }$ is the convex, $\sigma ((L^{\widehat{u}})^{\ast },L^{\widehat{u}})$%
--lower semicontinuous hull of $\alpha $. We want to show that $\pi ^{\ast
}=\alpha $. To this end, we prove that $\alpha $ is \emph{already} convex
and lower semicontinuous.

\begin{itemize}
\item[(a) ] $\alpha $ is convex: Let $Q(y)=yQ_{1}+(1-y)Q_{2}$ be the convex
combination of any couple of elements in $\mathcal{N}^{W}$ (if the $Q_{i}$
are not in $\mathcal{N}^{W}$ there is nothing to prove). Given any $\lambda
_{1},\lambda _{2}>0$ , define $\lambda (y)=\frac{1}{(1-y)\frac{1}{\lambda
_{2}}+y\frac{1}{\lambda _{1}}}$, so that $\frac{1}{\lambda (y)}=(1-y)\frac{1%
}{\lambda _{2}}+y\frac{1}{\lambda _{1}}$. Then 
\begin{equation*}
\begin{array}{c}
\alpha (Q(y))\leq \left[ x+\Vert Q^{s}(y)\Vert +\frac{E[\Phi (\lambda (y)%
\frac{dQ^{r}(y)}{dP})]-U_{0}^{W}(x)}{\lambda (y)}\right] \leq  \\ 
y\left[ x+\Vert Q_{1}^{s}\Vert +\frac{E[\Phi (\lambda _{1}\frac{dQ_{1}^{r}}{%
dP})]-U_{0}^{W}(x)}{\lambda _{1}}\right] +(1-y)\left[ x+\Vert Q_{2}^{s}\Vert
+\frac{E[\Phi (\lambda _{2}\frac{dQ_{2}^{r}}{dP})]-U_{0}^{W}(x)}{\lambda _{2}%
}\right]  \\ 
\end{array}%
\end{equation*}%
where the inequalities follow from the convexity of the norm and of the
function $(z,k)\rightarrow z\Phi (k/z)$ on $\mathbb{R}_{+}\times \mathbb{R}%
_{+}$, as already pointed out. Taking the infimum over $\lambda _{1}$ and $%
\lambda _{2}$ 
\begin{equation*}
\alpha (Q(y))\leq y\,\alpha (Q_{1})+(1-y)\,\alpha (Q_{2}).
\end{equation*}

\item[(b) ] $\alpha $ is lower semicontinuous: Since $\alpha $ is a convex
map on a Banach space, weak lower semicontinuity is equivalent to norm lower
semicontinuity. Suppose then that $Q_{k}$ is a sequence converging to $Q$
with respect to the Orlicz norm. We must prove that 
\begin{equation*}
\alpha (Q)\leq \liminf_{k}\alpha (Q_{k}):=L
\end{equation*}

We can assume $L=\liminf_{k}\alpha (Q_{k})<+\infty $, otherwise there is
nothing to prove. Now, it is not difficult to see that 
\begin{equation}
Q_{k}\overset{\Vert \cdot \Vert }{\rightarrow }Q\text{ iff }Q_{k}^{r}\overset%
{\Vert \cdot \Vert }{\rightarrow }Q^{r},Q_{k}^{s}\overset{\Vert \cdot \Vert }%
{\rightarrow }Q^{s}  \label{nh}
\end{equation}%
so that $Q_{k}^{r}\rightarrow Q^{r}$ in $L^{\widehat{\Phi }}$ and henceforth
in $L^{1}$. We can extract a subsequence, still denoted by $Q_{k}$ to
simplify notation, such that 
\begin{equation*}
\alpha (Q_{k})\rightarrow L\text{ and }Q_{k}^{r}\rightarrow Q^{r}\ a.s.
\end{equation*}%
So these $Q_{k}$ are (definitely) in $\mathcal{N}^{W}$, which is closed and
therefore the limit $Q\in \mathcal{N}^{W}$.

For all $k\in \mathbb{N}_{+}$ there exists $\lambda _{k}>0$ such that 
\begin{equation*}
\alpha (Q_{k})\leq x+\Vert Q_{k}^{s}\Vert +\frac{E[\Phi (\lambda _{k}\frac{%
dQ_{k}^{r}}{dP})]-U_{0}^{W}(x)}{\lambda _{k}}\leq \alpha (Q_{k})+\frac{1}{k}.
\end{equation*}

The next arguments rely on a couple of applications of Fatou Lemma to (a
subsequence of) the sequence $\left( \frac{\Phi (\lambda _{k}\frac{dQ_{k}^{r}%
}{dP})-U_{0}^{W}(x)}{\lambda _{k}}\right) _{k}$. Fatou Lemma is enabled here
by the condition $U_{0}^{W}(x)<u(+\infty )$ and by the convergence of the
regular parts $(\frac{dQ_{k}^{r}}{dP})_{k}$. In fact, one can always find an 
$\widetilde{x}$ such that $u(\widetilde{x})=U_{0}^{W}(x)$ and then the
Fenchel inequality gives the required control from below 
\begin{equation}
\frac{\Phi (\lambda _{k}\frac{dQ_{k}^{r}}{dP})-U_{0}^{W}(x)}{\lambda _{k}}+%
\frac{dQ_{k}^{r}}{dP}\widetilde{x}\,\geq 0.  \label{fatou}
\end{equation}%
The sequence $(\lambda _{k})_{k}$ cannot tend to $+\infty $. In fact, if $%
\lambda _{k}\rightarrow +\infty $, then a.s. we would have (remember that $%
\Phi $ is bounded below)%
\begin{eqnarray}
&&\liminf_{k}\frac{\Phi \left( \lambda _{k}\frac{dQ_{k}^{r}}{dP}\right)
-U_{0}^{W}(x)}{\lambda _{k}}=\liminf_{k}\frac{\Phi \left( \lambda _{k}\frac{%
dQ_{k}^{r}}{dP}\right) }{\lambda _{k}}  \notag \\
&\geq &\lim_{k}\frac{\left( \min_{y}\Phi (y)\right) }{\lambda _{k}}\mathbf{1}%
_{\{\frac{dQ_{k}^{r}}{dP}\wedge \frac{dQ^{r}}{dP}=0\}}+\lim_{k}\frac{\Phi
(\lambda _{k}\frac{dQ_{k}^{r}}{dP})}{\lambda _{k}}\mathbf{1}_{\{\frac{%
dQ_{k}^{r}}{dP}\wedge \frac{dQ^{r}}{dP}>0\}}  \notag \\
&=&\lim_{k}\frac{\Phi (\lambda _{k}\frac{dQ_{k}^{r}}{dP})}{\lambda _{k}\frac{%
dQ_{k}^{r}}{dP}}\frac{dQ_{k}^{r}}{dP}\mathbf{1}_{\{\frac{dQ_{k}^{r}}{dP}>0\}}%
\mathbf{1}_{\{\frac{dQ^{r}}{dP}>0\}}  \label{343}
\end{eqnarray}%
Since $\mathbf{1}_{\{\frac{dQ_{k}^{r}}{dP}>0\}}\mathbf{1}_{\{\frac{dQ^{r}}{dP%
}>0\}}\rightarrow \mathbf{1}_{\{\frac{dQ^{r}}{dP}>0\}}$ a.s. and, as already
checked, $\lim_{y\rightarrow +\infty }\frac{\Phi (y)}{y}=+\infty $ the limit
in (\ref{343}) is in fact $+\infty $ on the set $\{\frac{dQ^{r}}{dP}>0\}$
which has positive probability as $Q\in \mathcal{N}^{W}$. But then 
\begin{equation*}
\begin{array}{c}
L=\lim_{k}\{\alpha (Q_{k})+\frac{1}{k}\}=\lim_{k}\left\{ x+\Vert
Q_{k}^{s}\Vert +E\left[ \frac{\Phi (\lambda _{k}\frac{dQ_{k}^{r}}{dP}%
)-U_{0}^{W}(x)}{\lambda _{k}}\right] \right\} \\ 
\geq x+\Vert Q^{s}\Vert +E\left[ \liminf_{k}\frac{\Phi (\lambda _{k}\frac{%
dQ_{k}^{r}}{dP})-U_{0}^{W}(x)}{\lambda _{k}}\right] =+\infty \\ 
\end{array}%
\end{equation*}%
where in the inequality we apply \eqref{nh} and Fatou's Lemma.\newline
\indent Therefore there exists some compact subset of $\mathbb{R}_{+}$ that
contains $\lambda _{k}$ for infinitely many $k$'s, so that we can extract a
subsequence $\lambda _{k_{n}}\rightarrow \lambda ^{\ast }$. The inequality %
\eqref{fatou} ensures that $\lambda ^{\ast }$ must be strictly positive.
Otherwise, if $\lambda ^{\ast }=0$, the numerator of the fraction there
tends to $\Phi (0)-U_{0}^{W}(x)=u(+\infty )-U_{0}^{W}(x)>0$ and globally the
limit random variable would be $+\infty $. Finally, 
\begin{eqnarray*}
\alpha (Q) &\leq &x+\Vert Q^{s}\Vert +\frac{E[\Phi (\lambda ^{\ast }\frac{%
dQ^{r}}{dP})]-U_{0}^{W}(x)}{\lambda ^{\ast }} \\
&\leq &x+\liminf_{n}\left\{ \Vert Q_{k_{n}}^{s}\Vert +\frac{E[\Phi (\lambda
_{k_{n}}\frac{dQ_{k_{n}}^{r}}{dP})]-U_{0}^{W}(x)}{\lambda _{k_{n}}}\right\}
=L.
\end{eqnarray*}
\end{itemize}

Therefore, $\alpha =\pi ^{\ast }$ and the identity $\partial \pi (B)=%
\mathcal{Q}_{B}^{W}(x+\pi (B))$ in \eqref{subdiff} follows from %
\eqref{subdiffe} and \eqref{subdiffer}.

\item The bounds below are easily proved, 
\begin{equation}
\sup_{Q\in \mathcal{Q}_{0}^{W}(x)}Q(B)\leq \pi (B)\leq \sup_{Q\in \mathcal{\
N}^{W}}Q(B)  \label{bbbc}
\end{equation}%
since the first inequality follows from the fact that when $Q\in \mathcal{Q}%
_{0}^{W}(x)$, the penalty $\alpha (Q)=0$ (see \eqref{zero}) and the second
inequality holds because $\alpha $ is a penalty, i.e. $\alpha (Q)\geq 0$.
The first supremum is in fact a maximum, which is a consequence of
the\textquotedblleft Max Formula" as better explained in item 7 below. 
\newline
The case $W,B\in M^{\widehat{u}}$ is immediate from \eqref{bbbc} and from
the special form of the dual as stated in Corollary \ref{B}.

\item Let $\pi ^{\prime }(C,B)$ indicate the directional derivative of $\pi $
at $C$ along the direction $B$, i.e. $\pi ^{\prime }(C,B)=\lim_{b\downarrow
0}\frac{\pi (C+bB)-\pi (C)}{b}$. The so--called Max Formula (\cite[Theorem
4.2.7]{bz}) states that given a convex function $\pi $ and a continuity
point $C$, then 
\begin{equation*}
\pi ^{\prime }(C,B)=\max_{Q\in \partial \pi (C)}Q(B)
\end{equation*}%
So the first volume asymptotic becomes a trivial application of the Max
Formula with $C=0$, since $bB\in \mathcal{B}$ if $b\leq 1+\epsilon $ and 
\begin{equation*}
\lim_{b\downarrow 0}\frac{\pi (bB)}{b}=\pi ^{\prime }(0,B)=\max_{Q\in 
\mathcal{Q}_{0}^{W}(x)}\,Q(B),
\end{equation*}

because $\pi (0)=0$ and $\mathcal{Q}_{0}^{W}(x)=\partial \pi (0)$.\newline
For the second volume asymptotic, when $B\in M^{\widehat{u}}$ then $bB\in 
\mathcal{B}$ for all $b\in \mathbb{R}$. So, $\pi (bB)$ is well--defined and
for all $b>0$ we have that $\pi (bB)\leq \sup_{Q\in \mathcal{N}^{W}}Q(bB)$.
Therefore 
\begin{equation*}
\limsup_{b\rightarrow +\infty }\frac{\pi (bB)}{b}\leq \sup_{Q\in \mathcal{N}%
^{W}}Q(B).
\end{equation*}%
If we fix $Q\in \mathcal{N}^{W}$, the penalty $\alpha (Q)$ is finite and 
\begin{equation*}
\frac{\pi (bB)}{b}\geq Q(B)-\frac{\alpha (Q)}{b}\ \ \ \text{ for all }b>0
\end{equation*}%
so that 
\begin{equation*}
\liminf_{b\rightarrow +\infty }\frac{\pi (bB)}{b}\geq Q(B)\text{ for all }%
Q\in \mathcal{N}^{W}
\end{equation*}%
so that 
\begin{equation*}
\lim_{b\rightarrow +\infty }\frac{\pi (bB)}{b}=\sup_{Q\in \mathcal{N}%
^{W}}Q(B).
\end{equation*}%
Finally, the case $W,B\in M^{\widehat{u}}$ follows from the asymptotics just
proved and Corollary \ref{B}.

\item If $B$ and $-B$ are replicable with admissible strategies, then $%
Q(B)=c $ for all $Q\in \mathcal{N}^{W}$, whence in particular for the
\textquotedblleft zero penalty functionals" $Q\in \mathcal{Q}_{0}^{W}(x)$.
Therefore $\pi (B)=\max_{Q}\{Q(B)-\alpha (Q)\}=c$
\end{enumerate}
\end{proof}

\begin{remark}
As already noted in \cite[Remark 2.6]{bec}, if $B$ is not in $\mathcal{B}$
(e.g. a call option in a Black-Scholes model for an investor with
exponential preferences) but satisfies 
\begin{equation*}
B = B^* + \int_0^T H^*_sdS_s
\end{equation*}
where $B^* \in \mathcal{B}$ and the strategy $H^*$ is such that $\{ H+ H^*
\mid H \in \mathcal{H}^W\} = \mathcal{H}^W$, then one can apply the analysis
to $B^*$ and \emph{define } $\pi(B) = \pi(B^*) $.
\end{remark}


To better compare our results with the current literature, in the next
Corollary we specify the formula for $\pi$ in the exponential utility case.

\begin{corollary}
\label{CorH}Let $u(x)=-e^{-\gamma x}$, fix a loss variable $W$ and assume
that $\mathcal{N}^{W}\neq \emptyset $. If $B\in \mathcal{B}$ then: 
\begin{equation}
\pi _{\gamma }(B)=\max_{Q\in \mathcal{N}^{W}}\left[ Q(B)-\frac{1}{\gamma }%
\mathbb{H}(Q,P)\right] ,  \label{exp_price}
\end{equation}%
where the penalty term is given by: 
\begin{eqnarray}
\mathbb{H}(Q,P):= &&\gamma \Vert Q^{s}\Vert +H(Q^{r}|P)-U_{0}^{W}  \notag \\
&=&\gamma \Vert Q^{s}\Vert +H(Q^{r}|P)-\min_{Q\in \mathcal{N}^{W}}\left\{
H(Q^{r}|P)+\gamma \Vert Q^{s}\Vert \right\} .  \label{6authors_f}
\end{eqnarray}
\end{corollary}

Observe that, apart from the presence of the singular term $\Vert Q^{s}\Vert 
$, this result coincides with equation (5.6) of \cite{d6}. For a possible
interpretation of this term, both in \eqref{6authors_f} and in the general
representation \eqref{pi} let us define a \emph{catastrophic event} as a
random variable $\chi $ such that 
\begin{equation}
E[u(\chi )]>-\infty \mbox{   but   }E[u(\alpha \chi )]=-\infty \text{ for
some }\alpha >0.  \label{cat}
\end{equation}%
In other words, catastrophic events are given by random variables in the set 
\begin{equation}
\widehat{\mathcal{D}}:=\{f\in L^{\widehat{u}}\backslash M^{\widehat{u}}%
\mbox{
and }E[u(f)]>-\infty \}.
\end{equation}%
Since $Q^{s}$ vanishes on $M^{\widehat{u}}$, we conclude that 
\begin{equation}
\Vert Q^{s}\Vert =\sup_{f\in \mathcal{D}}Q^{s}(-f)=\sup_{f\in \widehat{%
\mathcal{D}}}Q^{s}(-f),  \label{gg}
\end{equation}%
so that the singular component is only relevant when computing $Q(f)$ for a
catastrophic $f\in \widehat{\mathcal{D}}$. Therefore, if $Q\in \mathcal{M}%
^{W}$ is a \textquotedblleft pricing measure" for which $\Vert Q^{s}\Vert >0$%
, then it might happen that $E_{Q^{r}}[f]>0$ for a catastrophic random
variable in the domain of optimization, despite the fact that $Q(f)\leq 0$
for all $f\in C^{W}$. Such $Q$ should then be used with caution. When
pricing the claim $B$ using the formula (\ref{exp_price}) or \eqref{pi}, the
pricing measures $Q\in \mathcal{M}^{W}$ that allow this unnatural behavior
are penalized with a penalization proportional to the relevance that $Q^{s}$
attributes to the catastrophic events according to \eqref{gg}.

\vspace{0.2in}

We conclude this section with some considerations on the risk measure
induced by $\pi$.

\begin{corollary}
\label{CorRho}Under the same hypotheses of Proposition \ref{propindiff} ,
the seller's indifference price $\pi $ defines a convex risk measure on $%
\mathcal{B}$, with the following representation: 
\begin{equation}
\rho (B)=\pi (-B)=\max_{Q\in \mathcal{N}^{W}}\{Q(-B)-\alpha (Q)\}.
\label{rhopi}
\end{equation}%
If both the loss control $W$ and the claim $B$ are in $M^{\widehat{u}}$,
then this risk measure has the Fatou property. In terms of $\pi $, this
means 
\begin{equation}
B_{n}\uparrow B\Rightarrow \pi (B_{n})\uparrow \pi (B)  \label{contbelow}
\end{equation}
\end{corollary}

\begin{proof}
The first part is a consequence of the above Proposition and the second part
follows from the fact that we have often stressed that when $W,B$ are in $M^{%
\widehat{u}}$ there is a version of the dual problem only with regular
elements $Q\in \mathcal{N}^{W}\cap L^{1}=\mathbb{M}_{\sigma }\cap \mathcal{L}%
_{\Phi }$. Consequently there is a representation $\rho (B)=\max_{Q\in 
\mathcal{N}^{W}\cap L^{1}}\{Q(-B)-\alpha (Q)\}$ on the order continuous
dual. But this implies the Fatou property (see e.g. \cite[Prop. 26]{BiaFri07}%
).
\end{proof}

\section{Comparison with existing literature}

The results above extend the literature on utility maximization with random
endowment when $u$ is finite on the entire real line. In fact, we allow the
semimartingale $S$ to be non locally bounded and as far as we know ours is
the first paper in this direction.\newline
\indent Also, the conditions we put on the claim $B$, that is $B\in \mathcal{%
A}_{u},$ are extremely weak - for the exponential utility $B\in \mathcal{A}%
_{u}$ simply means that $B$ satisfies (\ref{e}). The following list compares
our conditions on $B$ with those in the cited papers, \textit{which are all
formulated in the }$S$\textit{\ locally bounded case}.

To better compare these works, we stress that when $S$ is locally bounded,
we may select $W=1$ and therefore (see Corollary \ref{B}) the dual problem
can be formulated totally free of singular parts, as soon as $B\in M^{%
\widehat{u}}$, and we also get the representation of the optimal $f_{B}$ as
terminal value of an $S$-stochastic integral (Proposition \ref{hedge}).

\bigskip 

1. The first paper where a duality result of the type (\ref{1})-(\ref{2})
appeared - obviously with no singular components - is Bellini and Frittelli 
\cite{bef} Corollaries 2.2, 2.3, 2.4. In this paper, $u$ is finite on the
entire real line, $B$ is bounded, $W=1$, so that the admissible set of
trading strategies is $\mathcal{H}^{1}$ and $\mathcal{M}^{1}$ is the set of
local martingale measures.

2. The six Authors paper \cite{d6} (see also the related work by Kabanov and
Stricker \cite{kas}) considers only the exponential $u$. They extended the
results \cite{bef} in two respect. First they consider four different
classes of trading strategies (including $\mathcal{H}^{1}$) and secondly,
they assume condition \eqref{e} \textbf{plus} $B$ bounded from below. These
conditions clearly imply that $B\in \mathcal{B}=L^{\widehat{u}}\cap \mathcal{%
A}_{u}$.

3. Becherer's paper \cite{bec} also consider only the exponential case and
extend further the results in 1) and 2) above. His Assumption 2.4 
\begin{equation*}
E[e^{(\gamma +\varepsilon )B}]<+\infty ,\text{ }E[e^{-\varepsilon B}]<+\infty
\end{equation*}%
is however equivalent to saying that conditions \eqref{e} and (\ref%
{negative_bound}) hold, i.e. that $B\in \mathcal{B}=L^{\widehat{u}}\cap 
\mathcal{A}_{u}$.

4. For general utility $u$ finite on $\mathbb{R}$, the Assumption 1.6 on $B$
in Owen and Zitkovich \cite{oz} is on a different level, since it is a joint
condition on $B$ and the admissible strategies. This condition is not easy
to verify in practice, since it requires the prior knowledge of the dual
measures. Also, for economic reasons, we believe that it is better to state
the conditions on the claim only in terms of the compatibility with the
utility function.

\bigskip

\noindent \textbf{Acknowledgements.} S. Biagini warmly thanks P. Cheridito,
A. Hamel and B. Rudloff for some nice discussions on Risk Measures and
topics in Convex Analysis while visiting Princeton University.

\end{document}